\documentclass[aps,prx,twocolumn,showpacs,superscriptaddress,groupedaddress]{revtex4-1}
\usepackage{graphicx,subfigure,amssymb,amsmath,amsfonts,mathrsfs,color,braket,bm}
\usepackage[active]{srcltx}
\usepackage{times,txfonts}
\usepackage{hyperref}
\newcommand{\N}{\mathbb{N}}

\newcommand{\R}{\mathbb{R}}
\newcommand{\C}{\mathbb{C}}

\newcommand{\grp}[1]{\mathsf{#1}}
\newcommand{\spc}[1]{\mathcal{#1}}

\def\d{{\rm d}}

\def\>{\rangle}
\def\<{\langle}

\newcommand{\st}[1]{\mathbf{#1}}
\newcommand{\bs}[1]{\boldsymbol{#1}}

\newcommand{\map}[1]{\mathcal{#1}}
\newcommand{\Tr}{\operatorname{Tr}}

\newtheorem{theo}{Theorem}

\newtheorem{prop}{Proposition}

\newtheorem{defi}{Definition}

\def\qed{$\blacksquare$ \newline}

\begin{document}
\title
{Certifying quantumness: Benchmarks for the optimal processing \\ of generalized coherent and squeezed states}


\author{Yuxiang Yang}
\affiliation{Center for Quantum Information, Institute for Interdisciplinary Information Sciences, Tsinghua University, Beijing 100084, China}
\author{Giulio Chiribella}
\affiliation{Center for Quantum Information, Institute for Interdisciplinary Information Sciences, Tsinghua University, Beijing 100084, China}
\author{Gerardo Adesso}
\affiliation{School of Mathematical Sciences, The University of Nottingham, University Park, Nottingham NG7 2RD, United Kingdom}


\begin{abstract}

Quantum technology promises revolutionary advantages in information processing and transmission compared to classical technology; however, determining which specific resources are needed to surpass the capabilities of classical machines often remains  a nontrivial problem. To address such a problem, one first needs to establish the best classical solutions, which set benchmarks that must be beaten by any implementation claiming to harness quantum features for an enhanced performance. Here we introduce and develop a self-contained formalism to obtain the ultimate, generally probabilistic benchmarks for quantum information protocols including teleportation and approximate cloning, with arbitrary ensembles of input states generated by a group action, so-called Gilmore-Perelomov coherent states. This allows us to construct explicit fidelity thresholds for the transmission of multimode Gaussian and non-Gaussian states of continuous variable systems, as well as qubit and qudit pure states drawn according to nonuniform distributions on the Bloch hypersphere, which  accurately model the current laboratory facilities. The performance of deterministic classical procedures such as square-root measurement strategies is further compared with the optimal probabilistic benchmarks, and the state-of-the-art performance of experimental quantum implementations against our newly derived thresholds is discussed. This work provides a comprehensive collection of directly useful criteria for the reliable certification of quantum communication technologies.
\end{abstract}
\maketitle

\section{Introduction}

Quantum information technology \cite{nichu} is progressing at a fast pace. Theoretical and experimental results announcing groundbreaking demonstrations of the power of quantum hardware and quantum algorithms for the encoding and secure transmission of information receive media attention on an almost daily basis.  However, while basic building blocks for a distributed quantum communication network are available with current technology (encompassing e.g.~photonic architectures, solid-state memories, and hybrid interfaces thereof) \cite{QIkimble}, there remains as a crucial question to verify that the practical implementations of such devices are genuinely harnessing quantum resources and can provably outperform optimized special-purpose classical machines in realistic conditions of noise and decoherence. The problem of verifying quantum devices is in fact playing centre stage for the lively debate surrounding the operation of D-Wave `quantum' annealing processors, see e.g.~\cite{DWave}.

A general verification method consists in deriving tests, or {\it benchmarks}, which have to be passed by any realization of a quantum protocol, to corroborate its genuine use of quantumness; in this respect, benchmarks establish a quality control for realistic quantum information processing devices.
A standard way to construct such benchmarks is by focusing on a protocol and its figure of merit, and determine the corresponding classical threshold, i.e., the maximum value of the figure of merit which can be attained if the involved parties do not make use of any shared quantum resource, such as entanglement and quantum communication. Hence, if an actual experimental demonstration, despite being affected by unavoidable losses and imperfections, still manages to achieve a higher value of the figure of merit compared to the classical threshold, then the benchmark has been passed and the experiment is certified quantum, in the sense that its performance could not have been reached without the sharing and the utilization of suitable quantum resources.

The task of benchmarking quantum protocols has spurred an intense activity in the last two decades. The majority of relevant studies focused on the primal example of quantum teleportation and storage \cite{telep,brakim}, and derived benchmarks for the fidelity as a figure of merit \cite{bfkjmo}, for different classes of input states. To give an incomplete list, at present exact teleportation fidelity benchmarks are known in particular for input ensembles consisting of: pure qubits and qudits with uniform prior distribution \cite{benchqub,benchd}, pure displaced Gaussian states with finite-width Gaussian displacement distribution \cite{hammerer}, pure squeezed Gaussian states with unknown squeezing in one quadrature \cite{noi}, pure squeezed Gaussian states with finite-width squeezing distribution \cite{noi2}, pure displaced and squeezed Gaussian states with known squeezing and uniform displacement distribution \cite{owari}, pure displaced and squeezed Gaussian states with finite-width distributions of unknown squeezing and displacement \cite{noi2}. Other studies focused on numerical methods (e.g., based on semidefinite programming) to derive teleportation benchmarks for more  realistic sets of possibly mixed input states \cite{mariona}. Very recently, device-independent benchmarks for teleportation of qubit states  were investigated \cite{bancalnew}. Benchmarks applicable to other protocols have been developed as well. Benchmarks for non-unit-gain protocols, which include amplification, attenuation, and $N \rightarrow M$ quantum cloning of qubits, qudits and coherent states have been studied in detail \cite{benchd,sacchi,cerfprl,namikiprl,giulionew}; they reduce to the corresponding results for teleportation in particular limits (e.g. in the case of cloning for $N=M=1$).  Finally, an alternative approach to quantum benchmarks, testing the ability of a quantum device to transmit entanglement, has also been put forward \cite{norberto,killoran,imran}.

In this paper we develop a powerful and general formalism to obtain benchmarks for quantum protocols with arbitrarily distributed input ensembles generated by a group action, so-called generalized Gilmore-Perelomov coherent states (GPCS) \cite{gilmore,perelomov}. While this formalism allows us to straightforwardly recover a number of benchmark results existing in literature on teleportation, cloning and amplification of conventional coherent states (pure displaced Gaussian states) \cite{hammerer,namikiprl,giulionew}, it then goes significantly beyond by leading us to obtain novel analytical benchmarks for broad and relevant classes of input ensembles which can be used in present-day experiments. For example, our framework allows one consider  pure qubit and qudit states distributed according to realistically finite-width distributions on their respective Hilbert spaces, as well as single-mode and multi-mode Gaussian and non-Gaussian squeezed states with adjustable finite-width squeezing distributions, which are all examples of GPCS. Our machinery thus provides directly useful tests to certify the  quantum domain of current and future experiments encompassing discrete variable and continuous variable systems with realistic prior distributions.

The benchmarks we derive are probabilistic \cite{giulionew,noi2}: that is, we allow for the possibility to  discard some unfavourable trials. This increases the achievable fidelity threshold, at the expense of a nonunit success probability of the corresponding classical strategy. This procedure sets therefore the ultimate bounds that need to be surpassed by quantum implementations claiming to exploit authentic quantum resources to attain classically unreachable performances.
Probabilistic benchmarks are important not only as a stronger certificate of quantumness, but also because they are indispensable for  assessing  recent breakthrough  experiments in quantum optics,  implementing new quantum devices,  that achieve an enhanced  performance  at the price of a non-unit probability of success. The latest demonstrations of noiseless probabilistic amplifiers \cite{XiangRalph10,Ferreyrol10,usuga10,zavatta11} provide a fitting example of such devices.  Clearly, this new generation of quantum experiments   cannot be assessed using the old benchmarks, which referred only to deterministic strategies.
The increased interest in probabilistic devices provides also a strong motivation to analyze ensembles of input states with non-uniform priors, because the advantages of probabilistic protocols can only be seen in this setting \cite{XiangRalph10,Ferreyrol10,usuga10,zavatta11,giulionew,chiribellayang13}.

The paper is organized as follows.
In Section~\ref{SecGP} we develop the general machinery of probabilistic benchmarks for GPCS. In Section~\ref{sec:qudit} we showcase examples specialized to benchmarks for quantum cloning of qubit and qudit states with finite-width prior distributions. In Section~\ref{SecGau} we derive benchmarks for quantum cloning of relevant families of single-mode Gaussian states.  In Section~\ref{secPerelomov} we present benchmarks for general multimode non-Gaussian states belonging to the class of Perelomov squeezed states.
In Section~\ref{SecSQ} we investigate whether deterministic measure-and-prepare strategies (such as the square-root measurement) can achieve the bounds determined by the general probabilistic benchmarks of the previous sections.  In Section~\ref{SecTrans} we discuss the implementation of the optimal transposition of GPCS, a task which cannot be realized perfectly, and for which quantum protocols offer no advantage over the benchmarks. In Section~\ref{SecCC} we
conclude with a summary and an outlook of future theoretical and experimental directions. Some technical proofs and derivations are deferred to Appendices.

\bigskip



\section{Quantum benchmarks for the state transformation of generalized coherent states} \label{SecGP}

This section presents the general framework and the key result on  quantum benchmarks for generalized coherent states,  providing the foundation for the concrete applications worked out in the rest of the paper.
We  start from the general framework of quantum benchmarks and then specialize ourselves to the case of generalized coherent states, providing a general expression that allows one to evaluate the benchmarks explicitly.

\subsection{General framework for quantum benchmarks}\label{subsect:general}

\subsubsection{State transformation games}


A convenient  way to discuss quantum benchmarks  is in terms of  a game featuring two collaborating players, Alice and Bob, and a verifier, traditionally referred to as Victor \cite{bfkjmo}.
In this game, Victor presents Alice with an input state $\rho_x$,  with the label $x$ ranging in some set $\grp X$, and later he asks Bob to provide an output state, which should resemble as much as possible some desired target state $\rho_x'$.   In between,  Alice and Bob  team up to achieve the best possible approximation of the transformation $\rho_x \mapsto \rho_x'$, using the resources that are available to them.
 Here we assume that Alice and Bob know exactly the set of input states $\{\rho_x  ~|~  x\in \grp X\}$  and the set of target states $\{\rho'_x  ~|~  x\in \grp X\}$, but they do not know which particular input $\rho_x$ is presented to them.   We refer to games of this form  as \emph{state transformation games}.
 Most of the relevant examples of quantum information processing tasks can be cast in the form of state transformation games, by  suitably choosing the Hilbert spaces of the input (output) system---here denoted by $\spc H_{\rm in}$ ($\spc H_{\rm out}$)---and the input (output) states.  For example, choosing   $\spc  H_{\rm in}  =  \spc H_{\rm out}$ and $\rho'_x  =\rho_x$  one has the game of implementing  quantum teleportation and quantum memories for the set of states $\{\rho_x ~|~  x\in\grp X\}$.   For  $\spc H_{\rm in}  =  \spc H^{\otimes N},  \spc H_{\rm out}    =   \spc H^{\otimes M}, \rho_x=   \psi_x^{\otimes N}$ and $\rho_x'  =  \psi_x^{\otimes M}$ one has the game  of quantum cloning,  with an input of $N$ identical  copies  and an ideal  target of  $M$ output copies.

In a state transformation game, the role of the verifier Victor is to assess the quality of Alice's and Bob's implementation.   For this purpose, Victor  chooses the input state $\rho_x$ at random  with probability $p_x$ and    tests the output provided by Bob by performing a two-outcome measurement, described by a positive operator-valued measure (POVM) $\{T_x, \openone-  T_x\}$.  If the test gives the outcome corresponding to the operator $T_x$, then it is passed, otherwise, it  is failed.   Note that, if the target state is pure, say $\rho_x'  =  |\psi_x\>\<  \psi_x|$, then a natural choice for the operator $T_x$ is $  T_x  = |\psi_x\>\<\psi_x| $, so that the probability of passing the test is equal to the {\it fidelity} between Bob's output and  the target  state.
Nevertheless, we stress that our setting and our results are fully general, and include any operational test  $\{T_x, \openone-  T_x\}$ that the verifier may want to perform.     In the following we will consider the (average) probability of passing the test as our figure of merit, and denote it by $F$.


Suppose that Alice and Bob follow a deterministic protocol, described by a  trace-preserving completely positive (CP)  map  $\map M$.  In this case, the average
probability of passing Victor's test is
\begin{align}\label{detfig}
F_{\map M}  =    \sum_x    p_x     \Tr  [   T_x   \map M(\rho_x)] \, .
\end{align}
More generally, however,  the protocol adopted by Alice and Bob  can be probabilistic.  A probabilistic protocol will  either implement a desired quantum process---described by  a  trace non-increasing CP map $\map N$---or output a ``failure" message,  heralding the fact that the  process  $\map N$ did not take place.
 The probability that the process takes place on the input state $\rho_x$ is $p ({\rm yes}|x)   =   \Tr[  \map N(\rho_x)]$. When this happens,  the output state becomes
\begin{align}\label{rescaled}
\bar \rho_x   = \frac  {\map N(\rho_x)} {\Tr[\map N(\rho_x)]} \, .
\end{align}
The average probability that the process takes place is
\begin{align}\label{prob}  p_{{\rm yes}}  =  \sum_x   p_x  \Tr  [  \map N(\rho_x)] \, .
\end{align}
Conditional on the occurrence of  the process $\map N$, the figure of merit is given by
\begin{align}   F_{\map N}    =       \sum_x    p (x|  {\rm yes} )     \Tr  [   T_x     \bar  \rho_x]   \, ,
\end{align}
 where $p(x|{\rm yes})$ is the probability that the input state was $\rho_x$ conditioned on the information that $\map N$ took place.
Inserting Eq.~(\ref{rescaled}) into the above expression and using  Bayes' rule  $p(x|{\rm yes})=  p({\rm yes}|x) p_x/p_{{\rm yes}}$ one then obtains
\begin{align}\label{probfig}
F_{\map N}   =   \frac{\sum_x     p_x     \Tr[  T_x   \map N(\rho_x) ]    }{  \Tr  [ \map N( \rho)]} \, ,
\end{align}
where $\rho$ is the average input state
\begin{align}\label{rho}
\rho:  =  \sum_x  p_x  \rho_x  \, .
\end{align}
Note that when the device is deterministic (i.e.~when $\map N$ is trace-preserving) Eq.~(\ref{probfig}) coincides with Eq.~(\ref{detfig}).


\begin{figure}[t]
\centering
\includegraphics[width=8.2cm]{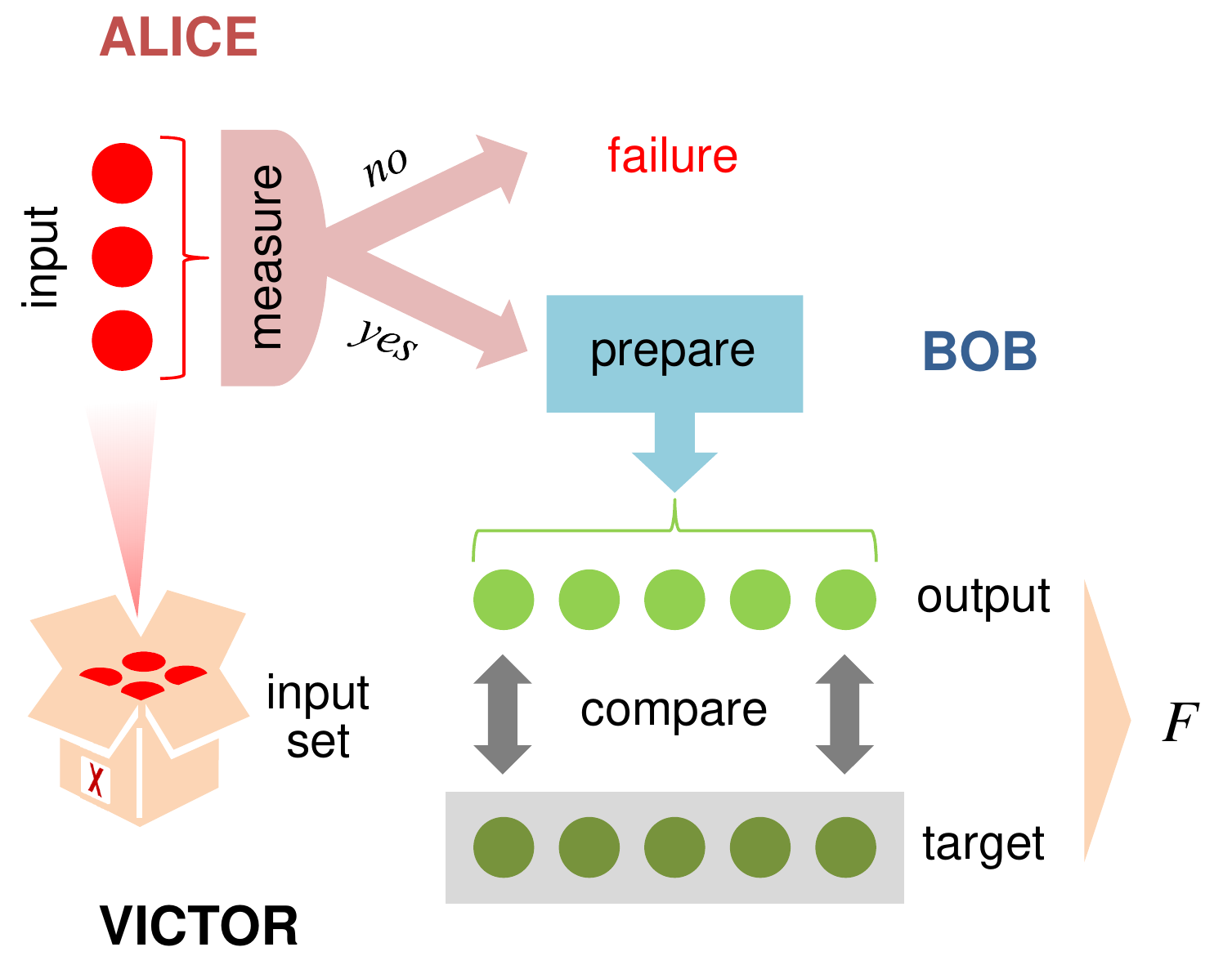}
\caption{(Color online) Scheme of a probabilistic measure-and-prepare (MP) state transformation game. The figure of merit $F$ is typically the fidelity between the output prepared by Bob and the target state set by the verifier Victor for a given input state supplied to Alice, averaged over the input set ${\mathsf X}$ according to a specified prior probability distribution. The maximum value $F_{\rm c}$ of $F$ over all MP protocols implementable by Alice and Bob sets a quantum benchmark for the transformation of the input ensemble of states. Any demonstration of state transformation of the input set  which achieves, on average, a figure of merit exceeding $F_{\rm c}$, implies necessarily that a quantum channel has been used as a resource  by Alice and Bob. }\label{fig:mp}
\end{figure}

\subsubsection{Ultimate performance achievable by  quantum strategies}
How well Alice and Bob can fare in a given state transformation game depends on which resources they are allowed to use.  In particular, if Alice and Bob are allowed to communicate quantum states  to one another (which implies, in particular, that they can share entanglement), they  can act effectively as a single super-player,  who in principle is able to implement  arbitrary quantum devices.  In this way, they can achieve the best performance allowed by  quantum mechanics,  given by the supremum of   the figure of merit in  Eq.~(\ref{probfig}) over all CP trace non-increasing maps.
This supremum, denoted by $F_q$, represents the ultimate limit that can be achieved by arbitrary processes in nature (assuming quantum mechanics as the correct theory of nature), including even  processes that take place with small probability.
 The actual value of $F_q$, which  can be computed  along the lines of Refs.~\cite{fiurasek,giulionew},  is given by
\begin{align}\label{qfid}
F_q  =    \left\|    \left[  \left (\openone_{\rm out}  \otimes    \rho^{-\frac 12}  \right) \Omega \left (\openone_{\rm out}  \otimes  \rho^{-\frac 12} \right)  \right]^{\Theta_{\rm in}}  \right\|_{\infty}
\end{align}
where $\openone_{\rm out}$ is the identity on the output Hilbert space, $\Theta_{\rm in}$ is the operation of partial transpose on the input Hilbert space,  the operator $\Omega$ is defined as
\begin{align}\label{Omega}
\Omega  :=  \sum_x  p_x  \,  T_x  \otimes \rho_x \, ,
\end{align}
and $\|  A\|_\infty$ denotes the operator norm, which for positive operators is given by the maximum expectation value
\begin{align}   \|A\|_\infty     =   \sup_{\|  \psi  \|  =  1   }  \<  \psi  | A |\psi\>  \, .
\end{align}
Clearly, when the set of states is continuous, the sums in Eqs.~(\ref{probfig}), (\ref{rho}) and (\ref{Omega}) have to be replaced by integrals and the probability distribution $p_x$ has to be replaced by a probability density $p(x) \d x$.


\subsubsection{Quantum benchmarks}

Suppose now that Alice and Bob are only allowed to communicate classical data. Under this restriction, the most general protocol that they can perform is   a \emph{measure-and-prepare (MP) protocol}, i.e.~a procedure where Alice measures  the input system and communicates the outcome to Bob, who uses this information to prepare the output system in a suitable quantum state.
The supremum of the figure of merit  of Eq. (\ref{probfig}) over all MP protocols is  the  \emph{quantum benchmark} for the task $\rho_x\to \rho_x'$.
By definition, if Alice and Bob can beat the quantum benchmark, they must have used {\it quantum} communication resources, either directly (with Alice sending quantum states to Bob) or indirectly (using the assistance of entanglement).


The most stringent benchmark is obtained if we allow Alice and Bob to use  probabilistic protocols, which provide an output to the verifier only if the measurement outcome  belongs to a subset of favourable outcomes.
 A generic MP protocol, as schematically depicted in Fig.~\ref{fig:mp}, is specified by a POVM  $\{P_y\}_{y\in\grp Y}$ and by a set of quantum states  $\{  \sigma_y\}_{y\in\grp Y_{\rm yes}}$ that are re-prepared when the measurement outcome belongs to the set of favourable outcomes  $\grp Y_{{\rm yes}}  \subseteq \mathsf Y$.  In the favourable instances, the protocol is described by the CP map $\widetilde{\map N}$ given by
 \begin{align}\label{MPchannel}
 \widetilde{ \map N}(\rho)  :  =  \sum_{y\in\grp Y_{\rm yes}}    \Tr  [ P_y  \rho ]  \,  \sigma_y \, .
 \end{align}

We denote by $F_{\rm c}$ the supremum of the figure of merit in Eq.~(\ref{probfig}) over  all probabilistic  MP protocols described by CP maps  $\widetilde {\map N}$ as in Eq. (\ref{MPchannel}).   Following the arguments of Ref. \cite{giulionew},  the actual value of $F_{\rm c}$ can be expressed as
\begin{align}\label{bench}
F_{\rm c}  =    \left\|  \left (\openone_{\rm out}  \otimes    \rho^{-\frac 12}  \right) \Omega \left (  \openone_{\rm out}  \otimes  \rho^{-\frac 12} \right)  \right\|_{\times}
\end{align}
where   $\Omega$ is the operator defined  in Eq. (\ref{Omega})
and $\|  A \|_\times$ denotes the  \emph{injective cross norm}  \cite{Banach}, which for positive operators is  given by the maximum expectation value on product states
\begin{align}\label{cross}   \|A\|_\times     =   \sup_{\|  \phi\|  =  \|  \psi\|    =  1   }  \<\phi|\<  \psi  | A |\phi\>  |\psi\>  \, .
\end{align}

Eq.~(\ref{bench}) quantifies the ultimate performance that can be achieved by arbitrary MP protocols, even allowing for protocols that produce an output with non-unit, arbitrarily small probability. Hence, if an experiment achieves a figure of merit which exceeds the classical threshold $F_{\rm c}$, then one can certify that the experimental setup has implemented a genuine quantum processing.   If each $T_x$ is of the form $T_x  =  |\psi_x\>\< \psi_x|$ for some pure state $|\psi_x\>$, then we call $F_{\rm c}$ the probabilistic \emph{classical fidelity threshold (CFT)} for the transformation $\rho_x \mapsto  |\psi_x\>\<  \psi_x|$.

Note the following important property:
\begin{prop}\label{prop} The CFT for the transformation  $\rho_x \mapsto  |\psi_x\>\<  \psi_x|$ is equal to the CFT for the three transformations
\begin{align*}
\rho_x   &\mapsto  | \psi^*_x\>\<  \psi^*_x|  \\
 \rho^*_x  &\mapsto  | \psi_x\>\< \psi_x|   \\
   \rho^*_x   &\mapsto  |\psi^*_x\>\< \psi^*_x|    \, ,
\end{align*} where $|\psi^*_x\>$  ($\rho^*_x$) is the complex conjugate of $|\psi_x\>$ ($\rho_x$) in a given basis.
\end{prop}
The validity of the above property can be immediately derived   from Eqs.~(\ref{Omega}), (\ref{bench}), and (\ref{cross}): Indeed, the value of the cross norm does not change if we replace the test operator $T_x$ and/or the state $\rho_x$ by its complex conjugate.  Similarly, the optimal MP protocol for one of the four tasks can be obtained from the optimal MP protocol of any of the others by applying complex conjugations where needed.  Note also that, at the density matrix level, complex conjugation is equivalent to transposition.

Combining this fact with Proposition \ref{prop}, one can see that the difference between the ultimate quantum fidelity $F_{\rm q}$ and the quantum benchmark $F_{\rm c}$ is nothing but the difference between two norms, namely the operator norm in Eq. (\ref{qfid}) and the cross norm in Eq. (\ref{bench}).

\subsection{Generalized coherent states} \label{secgccx}

We now specialize our analysis to the processing of generalized coherent states, providing an explicit and easily computable expression for the quantum benchmark  of Eq.~(\ref{bench}).
We consider  \emph{Gilmore-Perelomov coherent states (GPCS)}  \cite{gilmore,perelomov,perelomov2,ali}, a broad class of quantum states that contains the coherent states of the harmonic oscillator, the spin-coherent states  common in atomic physics, the families of coherent, squeezed, and pure Gaussian states in quantum optics, and (infinitely) many other families of non-Gaussian quantum states for continuous variable systems.

Technically, GPCS are associated to irreducible representations of locally compact groups   \cite{perelomov2,ali}.   Given a group $\grp G$,  a unitary projective representation $U:    g \mapsto  U_g$ acting on some Hilbert space $\spc H$, and a unit vector $|\phi\> \in  \spc H$,  we say that the states
\begin{align}\label{defgp}
|\phi_g\> :=   U_g  |\phi\>  \qquad g\in\grp G \,
\end{align}
are  GPCS    if the action of the representation  $U$  is irreducible in the subspace  spanned by them.

The best known example of GPCS are the coherent states of the harmonic oscillator, defined as
$$|\alpha\> :  =    D(\alpha)  |0\>  \qquad \alpha\in  \C$$ where $D(\alpha)$ is the displacement operator $D(\alpha)  =   \exp[  \alpha  a^\dag- \alpha^* a]$, $a$ and $a^\dag$ satisfy the commutation relation   $[a,a^\dag]=1$, and  $|0\>$ is the vacuum state, identified  by the equation $a|0\>  =  0$.
Another example is provided by the spin-coherent states
$$|  j,\theta,\phi\>  :  =  e^{i\frac {\theta} 2 \,  ( \sin \varphi    J_x  -   \cos \varphi   J_y  )} \,    | j,j\>  \qquad \theta\in  [0,\pi],  \varphi  \in  [0,2\pi] ,$$
where   $J_x,J_y,J_z$ are the angular momentum operators, satisfying the relations $[J_x,J_y]  =  i  J_z$, $[J_y,J_z]  =  i  J_x$,  $[J_z,J_x]  =  i  J_y$ and  $|j,j\>$ is the state identified by the equation $   J_z  |j,j\>  =   j  |  j,j\> $.

In addition to being GPCS, the harmonic oscillator coherent states  and the  spin-coherent states have an important feature: when one takes the tensor product of two states with the same label, one still obtains a GPCS.
For example, the product states   $ |j,\theta,\varphi\>  |j',\theta,\varphi\>$ are GPCS for every pair of values $(j,j')$, as they are unitarily equivalent to the spin-coherent state $|  j+j'  ,  \theta  , \varphi\>$.  Similarly the product states $  |g\alpha\>  |g'\alpha\>$ are GPCS for every pair of complex numbers $g$ and $g'$, as they are unitarily equivalent to the coherent state $  \left |   \sqrt{  |g|^2 +   | g' |^2} \alpha  \right \>$.
It turns out that this property is common to most of the examples that are relevant for applications, including squeezed states, pure Gaussian states, and arbitrary pure states of finite dimensional quantum systems.
The property plays a crucial role in our work and therefore it is convenient to give it  a name:
\begin{defi}
Two sets of GPCS $\{|\phi_{1,g}\>\}$ and $\{| \phi_{2,g}  \>\}$ are \emph{mutually coherent} if  the product states $\{  |\phi_{1,g}\>  |\psi_{2,g}\>\}$ are GPCS.
\end{defi}
The definition can be extended to more than two sets in the obvious way:   if  $\{  |\phi_{i,g}\>\}$ is a set of GPCS for every $i=  1,\dots, K$, we say that the $K$ sets  are mutually coherent if  the product states $\{  |\phi_{1,g}\>  |\phi_{2,g}\>  \cdots |\phi_{K,g}\>\}$ are GPCS.

In representation theory, the chief example of mutually coherent GPCS is the example of GPCS generated by the action of a semisimple Lie group on highest weight vectors \cite{ali}.  We recall that every semisimple Lie algebra can be decomposed into roots, and that highest (lowest) weight vectors are those that are annihilated by all the positive (negative) roots. For example, the Lie algebra of $\grp{SU(2)}$ can be decomposed as $su(2)  =   j_z  \oplus j_+  \oplus j_-$, where the $j_+$ is the positive root and $j_-$ is the negative root, and the vector $|j,j\>$ is a highest weight vector, as $j_+  |j,j\>  =  0$.
By definition,  if $|\phi_i\>\in\spc H_i$ is a highest weight vector for every $i$, then also $  |\phi_1\>  |\phi_2\>  \cdots |\phi_K\>$ is a highest weight vector and the subspace generated by the product vectors $|\phi_{1,g}\>   \cdots  |\phi_{K,g}\>$ is irreducible.  Hence, the states $\{|\phi_{1,g}\>   \cdots  |\phi_{K,g}\>\}$ are GPCS. Summarizing, GPCS that are generated from a highest weight vector are always mutually coherent.  This is the case of the spin-coherent states in atomic physics [associated to the group $\grp {SU} (2)$] and of most of the squeezed states considered in quantum optics [associated to the group $\grp{SU} (1,1)$].

\subsection{General quantum benchmark for GPCS}

We consider the task of transforming  $N$ copies of an input generalized coherent state $|\phi_g\>\in\spc H$ into $M$ copies of another, possibly different generalized coherent state $|\psi_g\>\in\spc K$.  As a figure of merit, we consider the maximization of the fidelity between the output state produced by the device and the target state $|\psi_g\>^{\otimes M}$ .
In the notation of Section \ref{subsect:general}, this means choosing  $x\equiv g$,  $\rho_g  =  \left(|\phi_g\>\<  \phi_g|  \right)^{\otimes N}$, $\rho_g'  =   \left(|\psi_g\>\<  \psi_g|  \right)^{\otimes M}$, and  $T_g  =     \left(|\psi_g\>\<  \psi_g|  \right)^{\otimes M}$.   Importantly, here we do not require the probability distribution $p(g)$ to be uniform. Instead, we allow it to be any probability distribution of the form
\begin{equation}\label{prior}
p_\gamma(g)  =d_{\gamma}|\<\phi_\gamma|\phi_{\gamma,g}\>|^{2}
\end{equation}
where  $|\phi_{\gamma,g}\>:=U_{\gamma,g}|\phi_\gamma\> $ is a GPCS in some Hilbert space $\spc H_\gamma$ and $d_\gamma$ is the normalization constant defined by
\begin{align*}
d_\gamma:=\left(\int\d g|\<\phi_\gamma|\phi_{\gamma,g}\>|^2\right)^{-1} \,,
\end{align*}
$\d g$ being the Haar measure on the group $\grp G$.
For example, when $\grp G$ is the group of translations in the plane, picking $ |\phi_{\gamma,g}\>$ to be the ordinary coherent states $  |\sqrt{\gamma }\alpha\>  =  D(\sqrt{\gamma}\alpha)  |0\>$, we have the Gaussian prior
\begin{align}\label{priogauss}
p_{\gamma}  ( \alpha)   &  =  \gamma   e^{  -   \gamma  |\alpha|^2} \, .
\end{align}
In general, the prior of Eq.~(\ref{prior}) has a precise operational meaning:  it is the prior that can be generated by preparing a GPCS $|\phi_\gamma\>$ and by measuring it via the coherent-state POVM
$$ P_{\gamma,g}  =  d_\gamma \, |\phi_{\gamma,g}\>\<\phi_{\gamma,g}| \, ,$$
whose normalization  $  \int \d g \,   P_{\gamma, g}    =  \openone$ follows from  Schur's lemma. Hence, we can imagine that the ensemble of input states $\{  |\phi_g\>,  p_{\gamma}  (g) \}$ is generated by performing the coherent-state POVM and, conditionally on outcome $g$, preparing the state  $|\phi_g\>$.
The uniform distribution, traditionally considered in most of the past literature, can be seen as a special case of  our probability distribution in Eq. (\ref{prior}),  obtained by choosing the space ${\spc H}_\gamma$ to be one-dimensional and   the group representation to be trivial ($U_g  \equiv 1$ for every $g\in\grp G$).

From now on, we will make the standing assumption  that the states $|\phi_g\>^{\otimes N}$, $|\psi_g\>^{\otimes M}$, and $|\phi_{\gamma,g}\>$ are mutually coherent GPCS.
Under this assumption, we provide a straightforward and in principle easy to compute expression for the quantum benchmark of Eq.~(\ref{bench}).  This expression is  the central result of the paper and is summarized in the following Theorem:
\begin{theo}\label{theo:bench}
 The probabilistic CFT for the transformation $|\phi_g\>^{\otimes N}\mapsto |\psi_g\>^{\otimes M}$    is given by:
\begin{align}\label{probCFT}
F_{\rm c}(\gamma)=\frac{\int\d g~ p_\gamma(g)~   |\<\psi|\psi_g\>|^{2M}  |\<\phi|\phi_g\>|^{2N}     }{\int\d g~ p_\gamma(g)~|\<\phi|\phi_g\>|^{2N}}.
\end{align}
\end{theo}
It is easy to see that there exists a probabilistic MP protocol (see Fig.~\ref{fig:mp}) that achieves the above fidelity.  The protocol consists in the following steps:
\begin{enumerate}
\item  measure the input states on the two-outcome POVM $\{P_{\rm yes},  P_{\rm no}\}$, where $P_{\rm yes} :  = ( |\phi\>\<  \phi|)^{\otimes N}$ and $P_{\rm no}  : =  \openone^{\otimes N}  -  P_{\rm yes}$;
\item   if the outcome is $\rm yes$, prepare the output state $|\psi\>^{\otimes M}$, if the outcome is $\rm no$, then declare ``failure".
\end{enumerate}
The successful instance of this protocol  is described by the quantum operation
$ \map N (\rho)  :   =     \Tr  [  P_{\rm yes}  \rho]  \, (|\psi\>\<\psi|)^{\otimes M}$
which takes place with average probability
\begin{align}\label{aveprob}   p_{\rm yes}   =  \int \d g  \,  p_\gamma  (g)   |\<\phi|\phi_g\>|^{2N}  \, .
\end{align}
Conditional on the occurrence of the successful outcome,  the  fidelity of the protocol is given by Eq.~(\ref{probfig}) and is exactly equal to $F_{\rm c}  (\gamma)$.   In the Appendix we prove that this value of the fidelity is optimal, i.e.~it is the highest value of the fidelity that can be achieved by arbitrary MP protocols. Therefore, we have that $F_{\rm c}(\gamma)$ is the CFT.

The result of Theorem \ref{theo:bench} can be extended to benchmarks where the performance is assessed on groups of $k$ output  systems.  Imagine that a quantum device attempts at performing the transformation $  |\phi_g\>^{\otimes N}  \to    |\psi_g\>^{\otimes M}$ and the performances of the device are tested according to the following recipe:
\begin{enumerate}
\item   pick a number $k\le M$  at random with probability $p(k)$;
\item  pick a $k$-element subset of the output systems, say $\grp S_{j}$, at random with probability $p (j|k)$;
\item test the fidelity between the state of the output systems in  $\grp S_j$  and $k$ perfect copies of the state $|\psi_g\>$.
\end{enumerate}
Even in this more general setting, the quantum benchmark has a simple form,  given by the CFT
\begin{align}\label{kcopy}
F_{\rm c} ( \gamma, p)   =   \sum_{k=1}^M  p(k)  \,    \frac{\int\d g~ p_\gamma(g)~   |\<\psi|\psi_g\>|^{2k}  |\<\phi|\phi_g\>|^{2N}     }{\int\d g~ p_\gamma(g)~|\<\phi|\phi_g\>|^{2N}}.
 \end{align}
The derivation of the benchmark is immediate:  On the one hand, it is clear that $F_{\rm c} (\gamma,p)$ is an upper bound, because for every fixed $k$ a generic MP protocol cannot perform better than the optimal MP protocol, whose fidelity is given by Eq.~(\ref{probCFT}) with $M  = k$.  On the other hand, it is easy to see that the MP protocol described after Theorem \ref{theo:bench} achieves probabilistically  the fidelity $F_{\rm c} (\gamma,p)$.

\section{Benchmarks for discrete variables}\label{sec:qudit}

Eqs. (\ref{probCFT}) and (\ref{kcopy}) cover a wide range of applications, including the benchmarks for teleportation and storage, complex conjugation, cloning and loss of generalized coherent states.  We  now analyze some of the most relevant cases and highlight the  applications of our benchmarks to several experimental setups.  For simplicity, we start  from discrete variables,
 while in the next two sections we will present results for continuous variables, both  in the Gaussian and non-Gaussian scenario.

\subsection{Pure qubit states}\label{sec:qubit}
Let us start from the simplest possible case:  benchmarking devices that transform $N$ equally prepared qubits into $M$ approximate copies. For $M=N$, the devices in question can be teleportation devices or quantum memories, where the state of the $N$ identically prepared qubits is stored.     For $M  >  N$ the devices are quantum copy machines (cloners), while for  $M  <  N$ they are noisy channels in which $N-M$ qubits are lost.  The case $M<N$ is interesting because it allows us to distinguish between different types of noise: on one hand, the noise resulting from the mere loss of particles, and on the other hand the noise due to a measurement, possibly by an eavesdropper controlling the transmission line.

Thanks to Eq.~(\ref{probCFT}), we can now evaluate  the CFT for all the above cases.
Let us denote by  $|\psi_{\theta,\varphi}\>^{\otimes N}$ the  state of $N$  identically prepared qubits, where
\begin{equation}\label{qbs}
|\psi_{\theta,\varphi}\> := \cos\frac{\theta}{2} \ket{0} + \sin\frac{\theta}{2}e^{i \varphi} \ket{1}  \qquad \theta \in [0,\pi] , \, \varphi \in [0, 2 \pi)\,,
\end{equation}
is the single-qubit state with Bloch vector  $\vec{r}_{\theta,\phi} = (\sin \theta \cos\varphi, \sin\theta \sin\varphi, \cos\theta)$.
 It is well known that the unitaries in $\grp{SU} (2)$ are represented by rotations on the Bloch sphere and that a generic pure state $|\psi_{\theta,\varphi}\>$ can be obtained by applying a rotation on the initial state $|\psi_{0,0}\>  \equiv |0\>$.
 Similarly, the states $\{  |\psi_{\theta,\varphi}\>^{\otimes N}\}$  are generated from the state $|0\>^{\otimes N}$ by the  $N$-fold tensor representation  of the group $\grp{SU(2)}$.  They are GPCS for every $N$, since they are  generated from the highest weight vector $  |0\>^{\otimes N}$.
  \begin{figure}[t]
\subfigure[]{
\includegraphics[height=3.4cm]{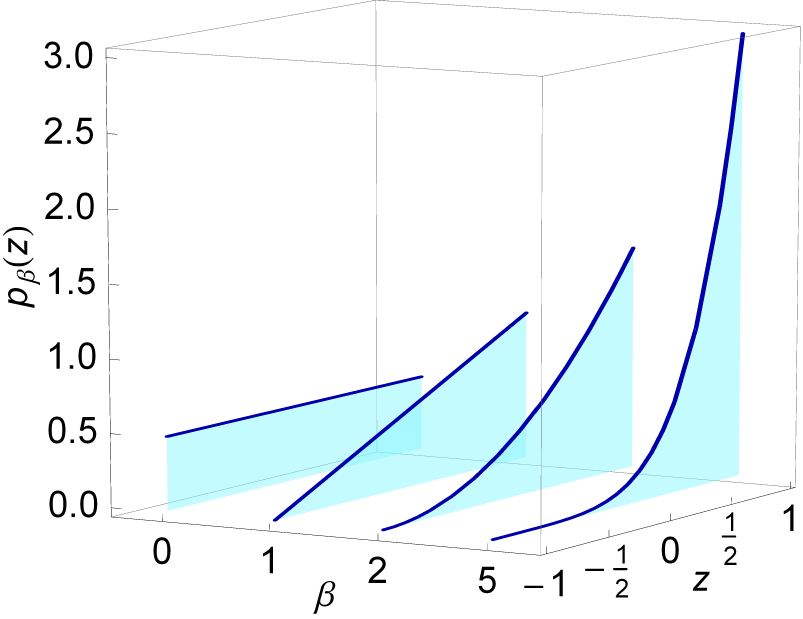}\label{pba}}\hspace*{.2cm}
\subfigure[]{
\includegraphics[height=3.4cm]{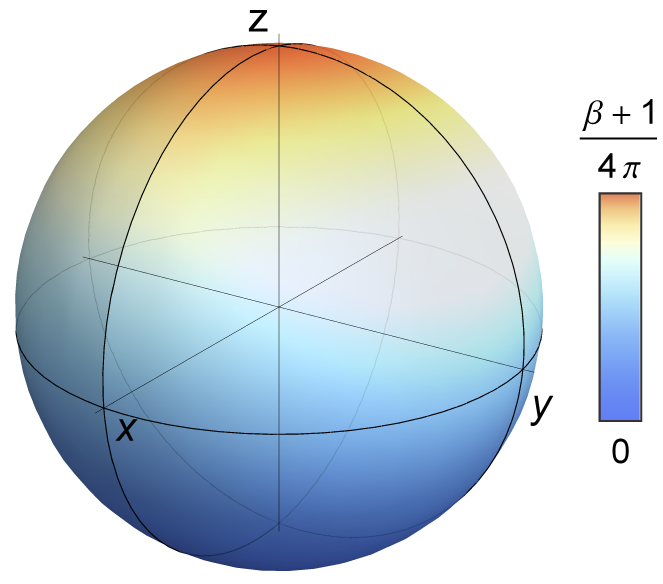}\label{pbb}}
\caption{(Color online) (a) Marginal prior probability distribution $p_\beta(z)$ for the $z$ component of the Bloch vector of single-qubit pure states, plotted for different values of the inverse width parameter $\beta$ ($\beta=0, 1, 2, 5$). (b) Overlay of the corresponding prior probability distribution $p_\beta(\theta,\varphi)$ for single-qubit states $|\psi_{\theta,\varphi}\>$ on the Bloch sphere (where $\theta,\varphi$ denote the polar and azimuthal angles), for $\beta=5$.}
\label{fig:qubitpriorall}
\end{figure}

We consider input states distributed according to the prior distribution
\begin{align}
\nonumber
p_\beta(\theta,\varphi)\d\theta\d\varphi& :=   d_\beta \left| \<   0 |\psi_{\theta,\varphi}\>\right|^{2\beta}  \,      \frac{\sin\theta\,  \d \theta\,   \d \varphi}{4\pi}    \\
 &  =  (\beta+1)\left(\cos\frac{\theta}{2}\right)^{2\beta+1}\sin\frac{\theta}{2}\d\theta~\frac{\d\varphi}{2\pi}  \, ,
\label{qubitprior}
\end{align}
which for integer $\beta$ is of the form Eq.~(\ref{prior}) [recall  that the $\grp{SU}(2)$-invariant measure on the sphere is given by $ \sin\theta \d \theta  \d \varphi/(4\pi)$].
Qubit states that are picked according to  the prior $p_\beta (\theta,\varphi)$   have a Bloch vector $\vec{r}_{\theta,\varphi}$ with totally random orientation in the $x$-$y$ plane and with  $z$-component distributed according to the marginal prior distribution $p_\beta(z) = 2^{-(\beta+1)}(\beta+1)(z+1)^\beta$, for $z \in [-1,1]$. Note that $p_\beta (z)$   becomes more peaked around $z=1$ as $\beta$ becomes larger, as illustrated in Fig.~\ref{fig:qubitpriorall} where we plot $p_\beta (z)$ for several values of $\beta$.

Now, using Eq.~(\ref{probCFT}) it is immediate to evaluate the CFT, which is given by
 \begin{align}
  F_{c}^{(2)}(\beta)
  &      =    \frac {N+\beta +1}{M+N+\beta +1} \, ,
     \label{eq:benchmarkqubit}
 \end{align}
where the superscript $(2)$ refers to the dimension of the input systems.  In Appendix~\ref{sec:proofqudit} we extend the validity of this formula from integer $\beta$ to arbitrary $\beta  >  0$, allowing one to interpolate continuously  from the completely flat distribution for $\beta=0$ to the Dirac-delta distribution for $\beta  \to \infty$.

Our Eq.~\eqref{eq:benchmarkqubit}   can be readily applied, for instance,  to validate implementations of quantum  telecloning of qubit inputs \cite{telecloning,gordontele}. The benchmark for these tasks is plotted in Fig.~\ref{fig:qudbench}(a)--(b).
The benchmark for probabilistic single-qubit teleportation, derived by setting the special case $N=M=1$ in Eq.~\eqref{eq:benchmarkqubit}, is
\begin{align}\label{qubittele}
F_{\rm c}^{(2){\rm tele}} (\beta)  =   \frac{\beta+2}{\beta+3} \, .
\end{align}

Interestingly, both Eqs.~\eqref{eq:benchmarkqubit}--\eqref{qubittele} reproduce known results in the literature \cite{benchqub,benchd} in the limit of uniform prior $\beta \rightarrow 0$.
Since these results had been derived for deterministic protocols, their coincidence with our benchmark shows that, for uniform prior,  probabilistic strategies offer no advantage: the absolute best MP protocol can be implemented with probability 1.  This feature is general:   the `magic' of probabilistic protocols for the processing of  GPCS fades away in the presence of maximal symmetry, as observed in~Ref.~\cite{chiribellayang13} (cf.~the Methods section therein and the discussion of the so-called ``many-world fairness"), and as proven explicitly in Appendix~\ref{app:optsquare} of this paper.


\subsection{Spin-coherent states}
The result for qubits can be extended in a straightforward way to spin-coherent states of spin-$j$ quantum systems with arbitrary $j$. From the mathematical point of view, the extension is trivial, because a system of spin $j$ can be thought as a composite system of $2j$ qubits. However, from the physical point of view, it is worth treating this case separately, since not all physical systems of spin $j$ are composite systems of qubits (think, for example, to the orbital angular momentum of an atom).

For spin systems we can consider a device that attempts at transforming $N$ copies of the coherent state $|j,j\>_{\theta,\varphi}$ into $M$ copies of the coherent state $|k,k\>_{\theta,\varphi}$.  Assuming a prior distribution of the form of Eq.~(\ref{qubitprior}), the  probabilistic CFT for this transformation is
\begin{align}\label{eq:benchmarkspin}
F_{\rm c}^{ (\rm{spin})}(\beta)=\frac{2jN+\beta+1}{2jN+2kM+\beta+1} \, ,
\end{align}
which follows from Eq.~(\ref{eq:benchmarkspin}) by making the substitutions $  N  \to  2j  N$ and $  M  \to  2k M$.

For $j=k$ and $M=N=1$,  Eq.~\eqref{eq:benchmarkspin} gives the probabilistic CFT for the teleportation and storage of spin-coherent states \cite{memorypolzik,telepolzik,bynes}, while for $M  >  N$  it provides the CFT for successful implementation of spin-coherent state cloning   \cite{demkowicz04}.   For $k>j$ and $M=N=1$, it gives the CFT for  ``spin-stretching", namely the task of enlarging the angular momentum of  quantum systems while preserving its orientation in space \cite{darianoperinotti09}.    To the best of our knowledge, the optimal spin stretching has not been implemented yet in atomic systems and beating the CFT in  Eq. (\ref{eq:benchmarkspin}) sets the goal for future experiments in this direction.


\subsection{Pure qudit states}

\begin{figure}[t]
\subfigure[]{
\includegraphics[height=3.3cm]{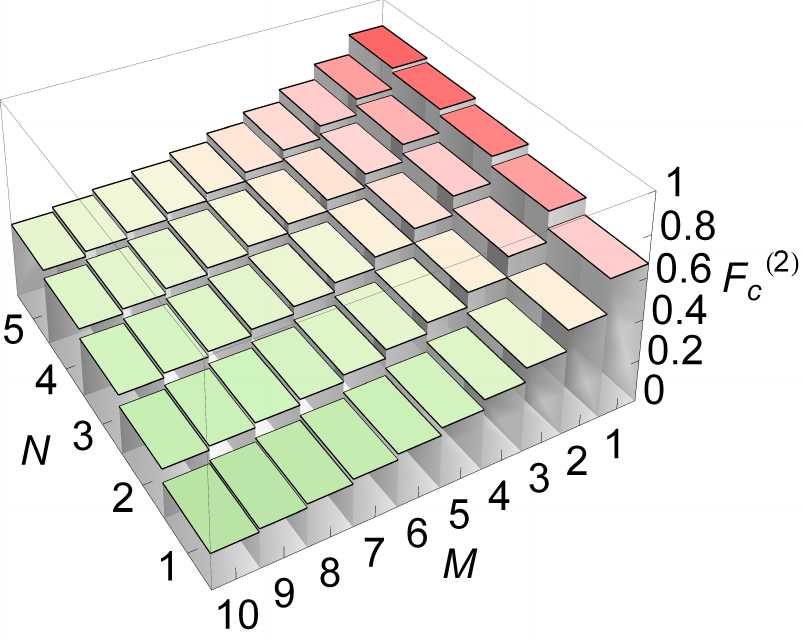}\label{q20}}\hspace*{.1cm}
\subfigure[]{
\includegraphics[height=3.3cm]{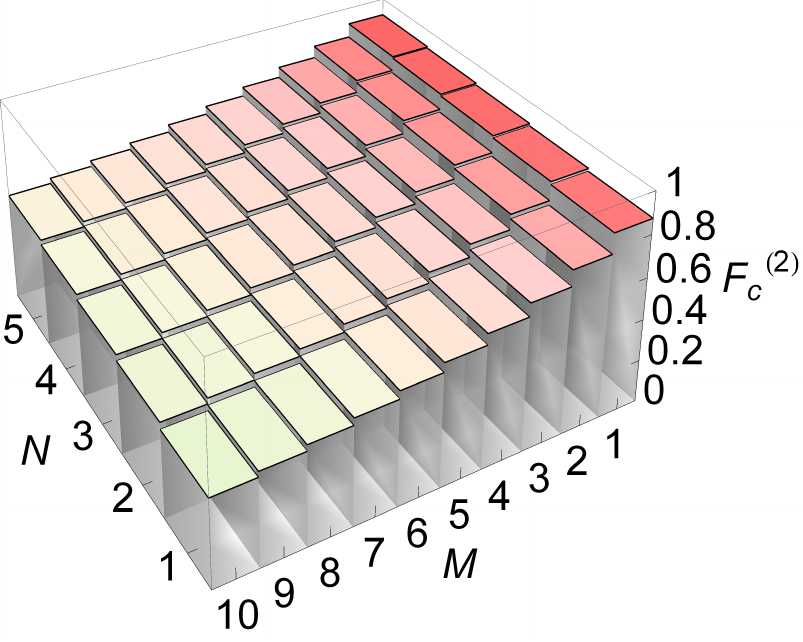}\label{q25}}\\
\subfigure[]{
\includegraphics[height=3.3cm]{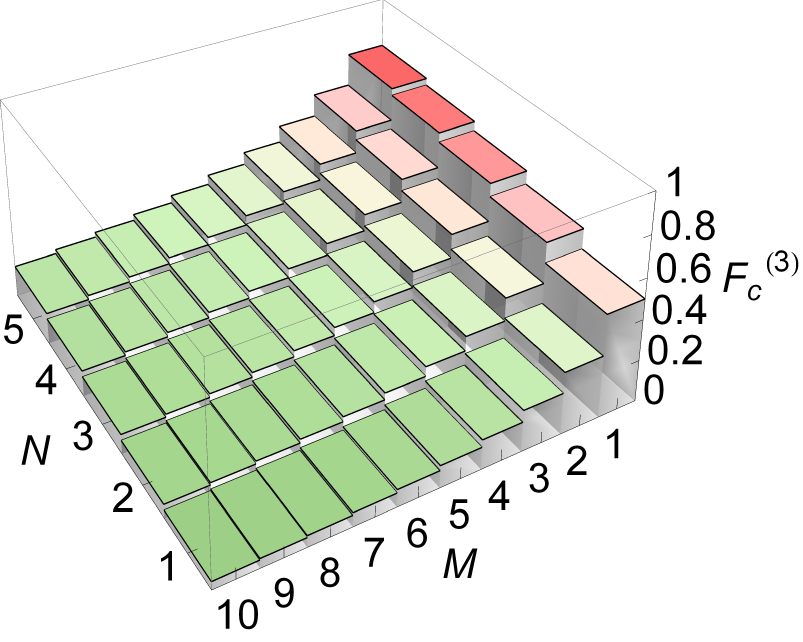}\label{q30}}\hspace*{.1cm}
\subfigure[]{
\includegraphics[height=3.3cm]{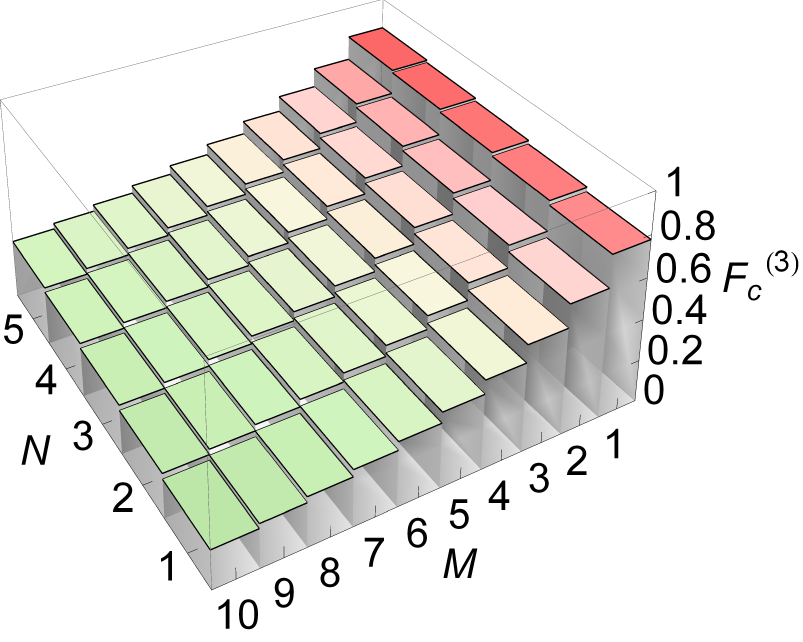}\label{q35}}\\
\subfigure[]{
\includegraphics[height=3.3cm]{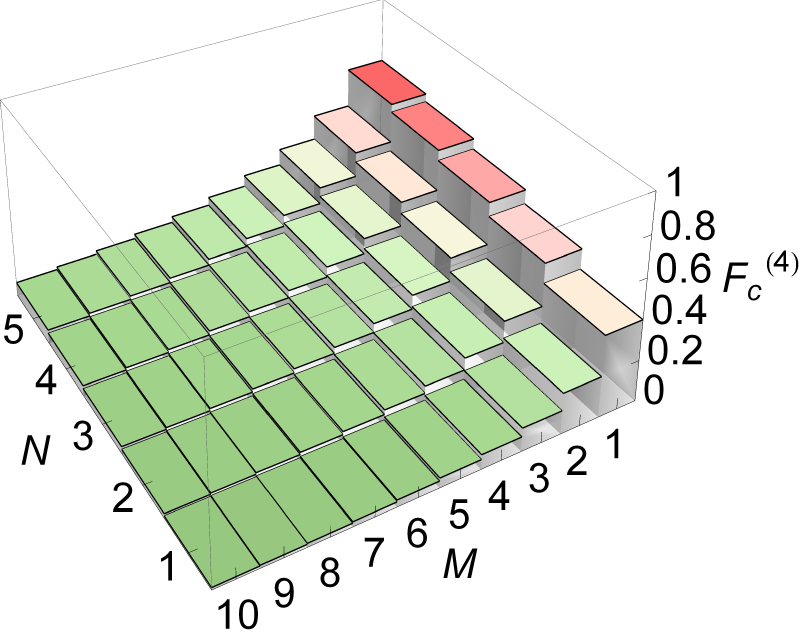}\label{p40}}\hspace*{.1cm}
\subfigure[]{
\includegraphics[height=3.3cm]{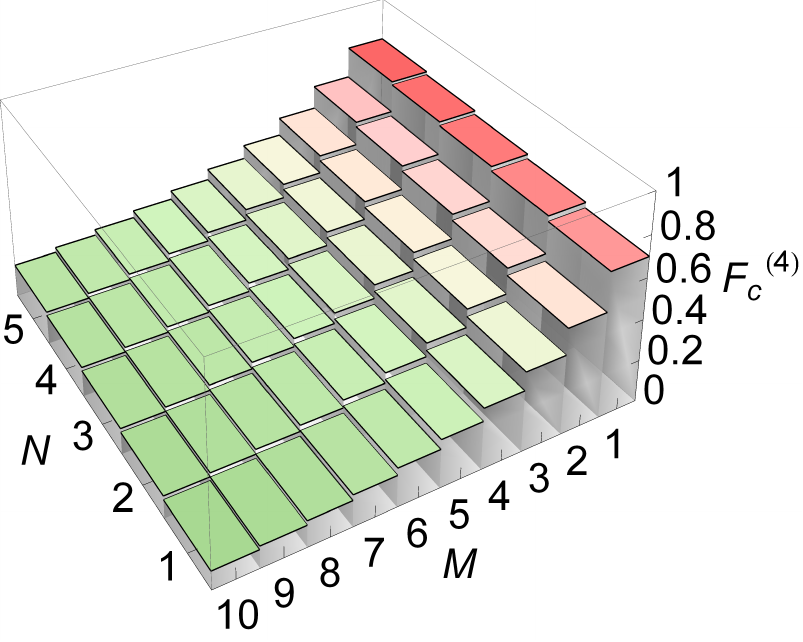}\label{p45}}\\
\caption{(Color online) Fidelity benchmark for the $N \rightarrow M$ transformation of arbitrary pure qudit states with (a)--(b) $d=2$, (c)--(d) $d=3$, and (e)--(f) $d=4$.  The input states are distributed according to a prior distribution $p_\beta$ dependent on an inverse width parameter $\beta$ as explained in the main text. In the first column [panels (a), (c), (e)], we set $\beta=0$, corresponding to a uniform distribution. In the second column [panels (b), (d), (f)], we set $\beta=5$, which gives a peaked distribution depicted in  Fig.~\ref{fig:qubitpriorall}(b) for the qubit case ($d=2$). The color legend for the CFT $F_{\rm c}^{(d)}$ in the bar charts is: $0$ \protect\includegraphics[width=1.3cm]{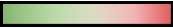} $1$.
\label{fig:qudbench}}
\end{figure}

We now extend our result for qubits to  arbitrary $d$-dimensional quantum systems (qudits), with $d<\infty$.  We analyze the task of transforming $N$ copies of a generic pure state  $|\psi\>\in\C^{d}$ into $M$ copies of the same state,  with a prior
\begin{align}\label{quditprior}
p_\beta(\psi)=d_\beta \, |\<0|\psi\>|^{2\beta},
\end{align}
where  $ |0\>$ is a fixed pure state and $d_\beta$ is the normalization constant defined by
\begin{align}\label{dbeta}
d_\beta  :=  \left(   \int  \d \psi  \,    |\<0|\psi\>|^{2\beta}  \right)^{-1}   \, ,
\end{align}
$\d \psi$ being the $\grp {SU} (d)$-invariant probability  distribution on the set of pure states. Again, for integer $\beta$ the prior is exactly of the form of Eq.~(\ref{prior}) and goes  from the uniform prior for $\beta=  0$ to a Dirac-delta centred  around the state $|0\>$ for $\beta \to \infty$.

The state $|\psi\>$ can be expressed in  the Hurwitz parametrization \cite{hurwitz,zyczkowski01} as
\begin{align}
|\psi\>   = \cos\theta_0  e^{i\varphi_0}|0\>   +  \sum_{j=1}^{d-1}  \cos  \theta_j   e^{i\varphi_{j}}   \left(  \prod_{n=0}^{j-1}   \sin\theta_{n} \right)  |j\>
\end{align}
where $\theta_j\in[0,\frac{\pi}{2})$ and $\varphi_j\in[0, 2\pi)$ for $n\in\{0,\dots,d-2\}$,  while $\varphi_{d-1}  = \theta_{d-1}  =  0$.  In this parametrization, the $\grp{SU}(d)$-invariant measure reads \cite{zyczkowski01}
\begin{align}\label{sudhaar}
\d \psi= \frac{(d-1)!}{\pi^{d-1}}  ~\prod_{j=0}^{d-2}\cos\theta_{j}(\sin\theta_{j})^{2 (d-j-1)-1}\d\theta_j\d\varphi_j \, .
\end{align}
In addition, one has $|\<  0|\psi\>  |  =  \cos \theta_0$ and the normalization constant in Eq.~(\ref{dbeta})   can be evaluated  explicitly as
\begin{align}\label{gammaint}
d_\beta=\frac{\Gamma(\beta+d)}{\Gamma(\beta+1)\Gamma(d)} \equiv{\beta+ d-1 \choose d-1} \, .
\end{align}
The CFT for the transformation $|\psi\>^{\otimes N}  \mapsto |\psi\>^{\otimes M}$, expressed as function of the dimension $d$ and the inverse width  $\beta$ of the prior, is then given by
\begin{align}\label{eq:benchmarkqudit}
F_{\rm c}^{(d)}(\beta)=\frac{{{N+\beta+ d-1}\choose{d-1}} }{{{M+N+\beta+ d-1}\choose{d-1}}},
\end{align}
and is plotted in Fig.~\ref{fig:qudbench}.

This expression follows directly from the combination of Eq.~(\ref{gammaint}) with Eqs. (\ref{probCFT}) and  (\ref{dbeta}), which give $F_{\rm c}^{(d)}  (\beta)  =  d_{N+\beta}/d_{M+N+\beta}$. Naturally, the quantum benchmark for qubits [Eq.~\eqref{eq:benchmarkqubit}] can be retrieved by setting $d=2$ in Eq.~\eqref{eq:benchmarkqudit}.
In Appendix  \ref{sec:proofqudit}  we extend the validity of Eq.~(\ref{eq:benchmarkqudit}) from integer $\beta$ to arbitrary positive real $\beta$.

 \section{Benchmarks for single-mode Gaussian states} \label{SecGau}
When moving from finite to infinite dimension, an almost trivial result is obtained if one takes the limit $d \rightarrow \infty$ in Eq.~ (\ref{eq:benchmarkqudit}): in this limit,  $F_{\rm c}^{(d)}$ tends to zero for every finite $\beta, M, N$. This reflects the fact that it is impossible for Alice and Bob to perform a reliable MP protocol simulating the transmission and processing of arbitrary pure states of  a continuous variable system.  Meaningful benchmarks need therefore to consider restricted sets of input states, the most interesting of which are those that can be prepared with the current experimental techniques.  An important example is provided by the Gaussian states, which are the principal resources for most quantum information protocols with continuous variable systems \cite{brareview,ourreview,book,pirandolareview,ournewreview}. Although they live in an infinite-dimensional Hilbert space, Gaussian states enjoy a simple mathematical description in terms of a finite set of degrees of freedom, given by the first and second moments of their creation and annihilation operators. Pure Gaussian states and many of their relevant subclasses are GPCS.  This is the case, e.g., of the conventional coherent states, generated by the Weyl-Heisenberg group, and of the squeezed states, generated by the group $\grp {SU}(1,1)$.

In the following we recall some recent results on quantum  benchmarks for $N \rightarrow M$ transformation of single-mode Gaussian states,  and we present several  new results on general non-Gaussian and multimode quantum states.

\subsection{Coherent states}

The coherent states of the harmonic oscillator are the prototype example of GPCS, and the optimal cloning \cite{CerfIpe2000,CerfIblisdir2000,Lindblad2000,CochraneRalph04,CerfKruger05} and optimal amplification \cite{RalphLund09,MarekFilip10,XiangRalph10,Ferreyrol10,usuga10,zavatta11,namikirapid,giulionew} of coherent states form a canonical chapter of continuous variable quantum information \cite{brareview,pirandolareview}. Recently,
the experimental breakthroughs in the demonstration of noiseless probabilistic amplifiers \cite{XiangRalph10,Ferreyrol10,usuga10,zavatta11} have sparked a renewed interest in the optimal processing of coherent states. Here the valuable feature is that quantum devices based on probabilistic filters can almost achieve the impossible tasks of amplification and cloning, with the catch that their probability of success decreases exponentially fast with the amplitude of the input coherent state. For this reason, being able to handle a non-uniform prior concentrated on the low-amplitude states is the key to discuss the new experiments of probabilistic amplification.      In this section we review the known benchmarks for the amplification of coherent states, showing how the general result of our paper provides an easy and quick derivation of the probabilistic CFT.

As mentioned in Section~\ref{secgccx}, a coherent state of a single-mode continuous variable system can be written as
\begin{equation}\label{coh}
\ket{\alpha} = D(\alpha) \ket{0}\,,
\end{equation}
where
\begin{equation}\label{displop}
D(\alpha)=\exp(\alpha  a^\dagger-\alpha^{\ast}  a)
 \end{equation}
 is the displacement operator,
 $ a$ and $ a^\dagger$ are  annihilation and creation operators obeying the commutation relation $[ a, a^\dagger]=1$, and $\ket{k}$ denotes the $k^{\text{th}}$ Fock state, with $\ket{0}$ being the vacuum. Now, suppose that an experimenter Alice is given $N$ copies of an input state $  |\alpha\>$  with complex displacement $\alpha$ distributed according to the Gaussian distribution
\begin{equation}\label{priorcoh}
p_\lambda(\alpha)=\frac{\lambda}{\pi} e^{-\lambda |\alpha|^2}
\end{equation}
with inverse width $\lambda^{-1}$, and suppose that her task is to produce $M$ copies of the target state $|g  \alpha\>$ for some fixed $g\in\C$.    When  $g$ is positive and larger than $1$, the state $ |g\alpha\>$ is an amplified version of $|\alpha\>$.  More generally,   allowing $g$ to be a complex number means allowing also for attenuation and phase shifting.     Teleportation, storage, complex conjugation, cloning, loss, amplification and attenuation of coherent states are all examples of the general task discussed here and can be retrieved by setting $N,M$ and $g$ to the appropriate values.

The study of quantum benchmarks for the processing of coherent states was first addressed in the case of teleportation by Hammerer \emph{et al} \cite{hammerer}, who proved the optimality of a fidelity threshold conjectured by Braunstein, Fuchs and Kimble  \cite{bfkjmo}. Later, benchmarks for deterministic amplification and attenuation were  put forward by Namiki \emph{et al} \cite{namikiprl}.   More recently, Chiribella and Xie derived the quantum benchmark for probabilistic amplification of coherent states \cite{giulionew}. Quite surprisingly, the value of the probabilistic benchmark coincides with the value of the deterministic benchmark obtained in Ref.~\cite{namikiprl}, implying that probabilistic MP protocols offer no advantage over deterministic ones in the case of coherent states.
   All these findings can be obtained from the general result of Theorem \ref{theo:bench}.  Indeed, since the  prior of Eq.~(\ref{priorcoh}) is of the form given in Eq.~(\ref{prior}), our general machinery applies and the probabilistic CFT can be obtained via  Eq.~(\ref{probCFT}). The calculation is immediate and gives the  result
   \begin{equation}\label{benchcoh}
F_{\rm c}^{(1c)}(\lambda) =  \frac{N+\lambda}{M  |g|^2   + N+\lambda}\,,
\end{equation}
where we use the suffix ``$(1c)$'' to indicate that the benchmark holds for  single-mode coherent states. For $N=M=1$, the above value coincides with the teleportation benchmark conjectured in \cite{bfkjmo} and proven in \cite{hammerer}, which converges to $\frac12$ for uniform prior distribution ($\lambda \rightarrow 0$).    For general $N$ and $M$, it reproduces the various bounds given in  \cite{cerfprl,namikiprl,giulionew}.



\subsection{Squeezed vacuum states}\label{Sec:SMSV}
An important class of single-mode squeezed states is generated by squeezing the vacuum.  Mathematically,  this means applying the squeezing operator
\begin{equation}\label{smsop}
S(\xi) = \exp\left[\frac12(\xi{ a^{\dagger}}{}^2-\xi^{\ast}  a^2)\right]
\end{equation}
to the vacuum state, thus obtaining the state
\begin{align*}
|\xi\>:=S(\xi)|0\> \, .
\end{align*}
In this Subsection we will provide the benchmark for states of this form, with arbitrary complex squeezing parameter $\xi$.
In the same spirit of the previous sections, we deal with the input state with a non-uniform prior, of the form $p_\beta(\xi)\mu(\d^2 \xi)$, where $p_\beta(\xi)$ is proportional to $|\<0 |\xi\>|^{2(\beta+2)}$ and  $\mu(\d^2 \xi)$ is the invariant measure on the squeezed states, given by
\begin{align*}
\mu(\d^2 \xi)=\sinh s\cosh s\d s\d\theta/(2\pi)  \qquad \xi := s e^{i \theta}
\end{align*}
(see e.g. \cite{chiribellasu11}).  Precisely, the prior is given by
\begin{align}\label{priorperelomov}
\nonumber
p_\beta(\xi)\mu(\d^2 \xi)&:=d_\beta |\<\xi|0\>|^{2(\beta+2)}\mu(\d^2 \xi)\\
&=\frac{\beta\sinh s~\d s}{(\cosh s)^{\beta+1}}~\frac{\d\theta}{2\pi}
\, .
\end{align}
The single-mode squeezed vacuum is a special case of a much broader category of (generally multimode and generally non-Gaussian) squeezed states, known as \emph{Perelomov squeezed states} \cite{perelomov2}.
For this category of states the quantum benchmark can be derived in a unified way, which will be presented in Section~\ref{secPerelomov}.     We anticipate here  the probabilistic benchmark for the single-mode squeezed vacuum, given by
\begin{align}\label{benchSMSV}
F_{c}^{(1s)}(\beta)=\frac{N+\beta}{M+N+\beta}.
\end{align}
Note that for $N=M=1$ the benchmark \eqref{benchSMSV} reduces to the benchmark for teleportation of single-mode squeezed vacuum states recently  obtained by some of us in  Eq.~(3a) of Ref.~\cite{noi2}.

The single-mode squeezed vacuum states are frequently used as an approximation of  the so-called ``even Schr{\"o}dinger cat states'' \cite{cat2,cat1,gran2007,telecat}, which are defined as superpositions of two coherent states
\begin{align*}
|\psi_{\rm even\ cat}(\alpha)\>:=\frac {|\alpha\>+|-\alpha\>}{\sqrt{2\left(1+e^{-2|\alpha|^2}\right)}} \, .
\end{align*}
The approximation is accurate when the squeezing degree  $|\xi|$ is small, and rapidly worsens as $|\xi|$ increases.
Choosing a prior distribution that is concentrated around $\xi  = 0$  (i.e.~choosing a sufficiently large $\beta$), we can guarantee that the input squeezed state will be with high probability a good approximation of an even cat state  and we can use  Eq. (\ref{benchSMSV}) as a teleportation and cloning validation criterion for  even cat input states.   Needless to say, the feature that our benchmarks hold for generally  non-uniform prior is essential here.


\subsection{Arbitrary single-mode Gaussian states}

A general pure single-mode Gaussian state can be written as
\begin{equation}\label{gauss}
\ket{\alpha,\xi}
:= D(\alpha) S(\xi) \ket{0}\,,
\end{equation}
where $D(\alpha)$ is defined in Eq.~(\ref{displop}) and $S(\xi)$ is defined in Eq.~\eqref{smsop}. Pure single-mode Gaussian states are thus entirely specified by their displacement vector $\alpha \in \mathbb{C}$, their squeezing degree $s\in \mathbb{R}^+$, and their squeezing phase $\theta \in [0,2\pi)$.   Quite conveniently for our purposes, they are also a family of GPCS, generated by the action of the Jacobi group on the vacuum \cite{Jacobi}.

\begin{figure}[t]
\subfigure[]{
\includegraphics[height=3.5cm]{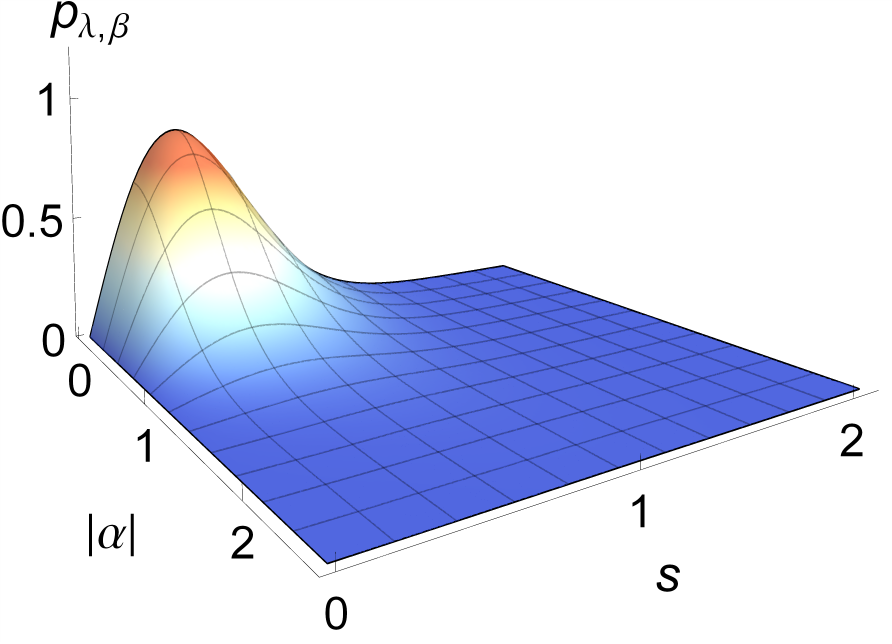}\label{gpba}}\hspace*{.2cm}
\subfigure[]{
\includegraphics[height=3.5cm]{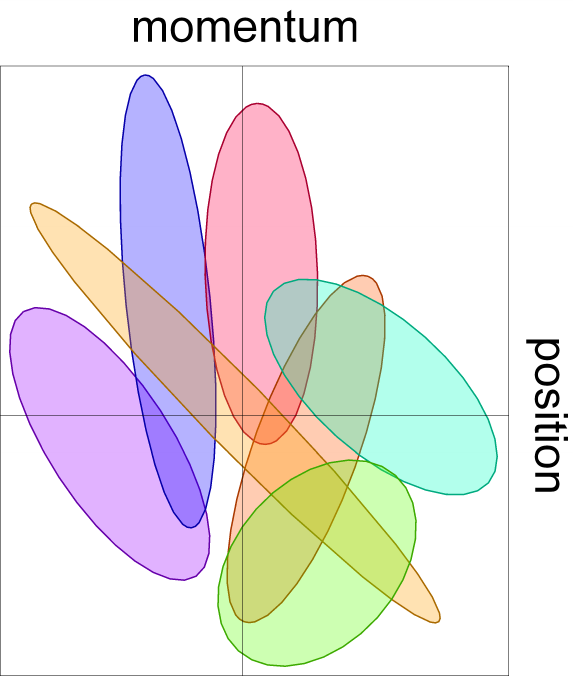}\label{gpbb}}
\caption{(Color online) Prior probability distribution over the input ensemble of pure single-mode Gaussian states, Eq.~(\ref{plambdabeta}), setting the inverse width parameters to $\lambda=2$ and $\beta=6$. Panel (a) depicts the marginal prior distribution for displacement $\alpha$ and squeezing degree $s$ after integrating Eq.~(\ref{plambdabeta}) over the squeezing phase $\theta$, yielding  $p_{\lambda,\beta}(\alpha,s) =  \pi^{-1}  \lambda \beta  e^{-\lambda |\alpha|^2}  \sinh s (\cosh s)^{-\beta    -2} I_0\big[\lambda|\alpha|^2   \tanh s\big]$ \cite{noi2}, where $I_0$ is a modified Bessel function. Panel (b) depicts cross-sections of the phase-space Wigner functions for a small sample of pure single-mode Gaussian states with parameters $\alpha$, $s$, and $\theta$ randomly sampled according to the considered prior distribution. The correspondence between the parameters of a Gaussian state and the form factor of the corresponding cross-section (which is an ellipse obtained by cutting the two-dimensional Gaussian Wigner function) is as follows: the centre of the ellipse has phase-space coordinates $(q, p) \equiv (\sqrt{2} {\rm Re} \alpha, \, \sqrt{2} {\rm Im} \alpha)$, the ellipse is rotated by $\theta/2$ with respect to the horizontal axis, and the ratio between the lengths of the semiaxes is $e^{2s}$; see e.g.~\cite{ournewreview} for more detail.}
\label{fig:priorgauss}
\end{figure}

Following the general prescription of Eq.~(\ref{prior}), we consider the following prior distribution \cite{noi2}
\begin{eqnarray}
\hspace*{-1cm}p_{\lambda,\beta}(\alpha,s,\theta)&=&\frac{\lambda \beta}{2 \pi^2}        \frac{ e^{-  \lambda|\alpha|^2   + \lambda  {\rm Re} (  e^{-i\theta} \alpha^2)  \tanh s    }   \sinh s}{(\cosh s)^{\beta + 2}}  \,.\label{plambdabeta}
\end{eqnarray}
This prior, depicted in Fig.~\ref{fig:priorgauss}, depends on two inverse width parameters, $\lambda$ regulating the distribution of the displacement $\alpha$, and $\beta$ regulating the distribution of the squeezing degree $s$, while the squeezing phase $\theta$ is uniformly distributed.
Note that the prior in (\ref{plambdabeta}) can be written as $p_{\lambda,\beta}  (\alpha,\xi)  \propto   |\<  0|  \lambda \alpha, \xi\>|^2   |\<  0  |  \xi\>|^{2(4+\beta)}  \nu(\d^2\alpha,\d^2 \xi) $ where $  \nu(\d^2\alpha,\d^2 \xi)  =  \d^2\alpha  \sinh s (\cosh s)^3 \d s \d \theta$ is the invariant measure under the joint action of displacement and squeezing  \cite{perelomov2}.
In the case of no squeezing  ($\beta\to\infty$), this prior correctly reproduces the Gaussian prior for coherent states, with    $\lim_{\beta \rightarrow \infty} \int \d^2\xi\ p_{\lambda,\beta}(\alpha,\xi) = p_\lambda(\alpha)$. Similarly, the marginal prior  of Eq.~(\ref{priorperelomov}) for squeezed states is recovered by integrating $p_{\lambda,\beta}(\alpha,\xi)$ over $\alpha$, namely $\int \d^2\alpha\ p_{\lambda,\beta}(\alpha,\xi) = p_\beta(\xi)$ \cite{noi2}.


For integer values of $\lambda, \beta$, the benchmark for $N \rightarrow M$ transformation of arbitrary pure single-mode Gaussian states distributed according to the prior in (\ref{plambdabeta}) is then given by evaluating Eq.~(\ref{probCFT}), which results in the following CFT, where the suffix ``(1cs)'' stands for one-mode coherent (i.e.~displaced) squeezed states:
\begin{eqnarray}
F_{\rm c}^{(1cs)} (\lambda,\beta)
&=&  \frac{(N+\lambda)(N+\beta)}{(N+M+\lambda)(N+M+\beta)} \,.
\label{bench1mg}
\end{eqnarray}

\begin{figure}[t]
\subfigure[]{
\includegraphics[height=3.3cm]{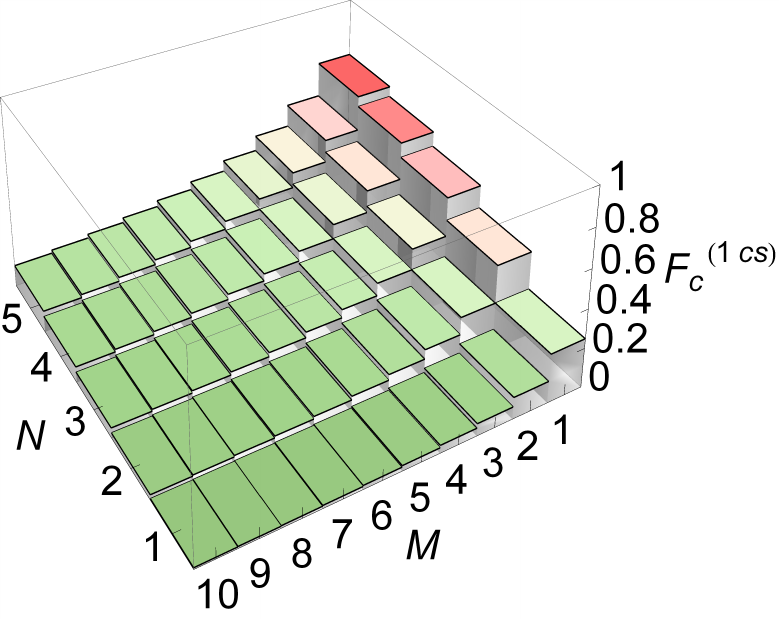}\label{g00}}\hspace*{.1cm}
\subfigure[]{
\includegraphics[height=3.3cm]{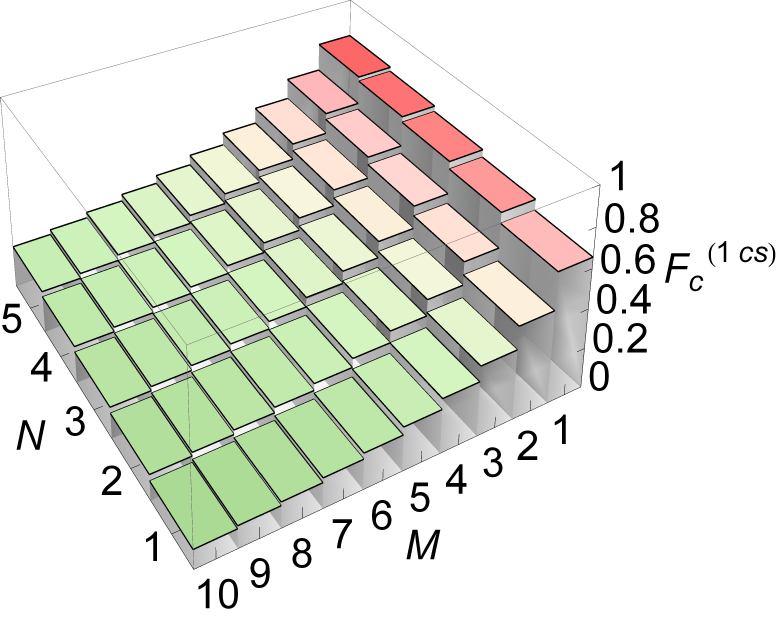}\label{g26}}\\
\caption{(Color online) Fidelity benchmark for the $N \rightarrow M$ transformation of arbitrary pure single-mode Gaussian states, distributed according to the prior $p_{\lambda,\beta}$ of Eq.~(\ref{plambdabeta}), with (a) $\lambda = \beta = 0$ (input ensemble of infinite energy), and (b) $\lambda = 2$, $\beta = 6$ (peaked distribution as in Fig.~\ref{fig:priorgauss}). The color legend for the CFT $F_{\rm c}^{(1cs)}$ in the bar charts is: $0$ \protect\includegraphics[width=1.3cm]{fclegendf.pdf} $1$.
\label{fig:benchgg}}
\end{figure}

The benchmark of Eq.~(\ref{bench1mg}) is plotted in Fig.~\ref{fig:benchgg}.
 For $N=M=1$, this result reproduces the very recent benchmark for teleportation and storage of pure single-mode Gaussian states, obtained by some of us in Ref.~\cite{noi2}. Interestingly, the benchmark for pure Gaussian states is equal to the product of the benchmarks for coherent states [Eq.~(\ref{benchcoh}) with $g=1$] and the benchmark for squeezed vacuum states [Eq. \eqref{benchSMSV}]. These two benchmarks are retrieved in the limits $\beta\to\infty$ (no squeezing) and $\lambda \to\infty$ (no displacement), respectively.
 One can numerically verify that the benchmark in Eq.~(\ref{bench1mg}) holds for generally noninteger values of $\beta$ and $\lambda$ as well, by using a laborious but straightforward operator approach \cite{noi2}.

\section{Benchmarks for non-Gaussian multimode squeezed states}\label{secPerelomov}
In the following, we turn our attention from single-mode Gaussian states to multimode non-Gaussian states. Specifically, we  introduce the benchmark for a family of squeezed states which were first proposed by Perelomov \cite{perelomov,perelomov2}.   The family of Perelomov squeezed states includes Gaussian states like the single-mode squeezed vacuum states (discussed in Subsec.~\ref{Sec:SMSV}) and, more intriguingly, also several non-Gaussian states. For example, the state obtained by squeezing a single-photon Fock state belongs to this class; such a state is a good approximation of an ``odd Schr{\"o}dinger cat state" and has  applications in hybrid continuous variable quantum information processing \cite{cat2,cat1,gran2007,telecat,hybridrev}.

\subsection{Benchmark for general Perelomov squeezed states}
Perelomov squeezed states are characterized  by an index  $j>0$ and by a squeezing parameter $\xi  =   s e^{i\theta}$.    For every fixed value of $j$ one has a family of squeezed states, generated by the action of a suitable squeezing operator on a given state.  The members of the family are states of the form
\begin{align}\label{eq:perelomovstate}
\left|\xi, j\right\>&:=\frac{1}{(\cosh s)^{j}}\sum_{n=0}^{\infty}\sqrt{{n+j-1\choose n}}(e^{i\theta}\tanh s)^{n}\left|\psi_{n}^{(j)}\right\> \,  ,
\end{align}
where $\big\{\big|\psi_{n}^{(j)}\big\>\big\}_{n\in\N}$ is a given orthonormal  basis.  For a fixed value of $j$, we call the states in this family the \emph{Perelomov-$j$ states}. Many known families of  squeezed states can be expressed as Perelomov-$j$ states.
 For example, the single-mode squeezed vacuum states, discussed in Subsec.~\ref{Sec:SMSV}),  can be viewed as Perelomov-$\frac12$ states by setting  $|\psi_{n}^{(j= 1/2)}\big\>  \equiv  |2n\>$ in Eq. (\ref{eq:perelomovstate}).    Similarly, the squeezed single-photon states can be viewed as  Perelomov-$\frac32$ states by setting $|\psi_{n}^{(j= 3/2)}\big\>  \equiv  |2n+1\>$.

Let us consider the transformation $\left|\xi ,j \right\>^{\otimes N}\to \left|\xi , k\right\>^{\otimes M}$, where  $N$ copies of a Perelomov-$j$ state are mapped into  $M$ copies of the corresponding Perelomov-$k$  state.  Among the possible examples one has not only state teleportation and  cloning, but also more exotic tasks, such as the  transformation of squeezed vacuum states into squeezed single-photon states.
Assuming  the same prior used in  Eq.~\eqref{priorperelomov} it is easy to evaluate the benchmark from Eq. (\ref{bench}).  The calculation, presented in Appendix~\ref{SecProofMultimode}, yields the following    CFT for Perelomov squeezed states
\begin{align}\label{eq:benchmarkperelomov}
F_{c}^{({\rm P})}(\beta)=\frac{2jN+\beta}{2kM+2jN+\beta}.
\end{align}

Plenty of appealing results can be derived from Eq.~\eqref{eq:benchmarkperelomov}, e.g.~the benchmark for the single-mode squeezed states in Eq.~\eqref{benchSMSV}. In the rest of this section
we will show a selection of what we believe are the the most meaningful ones.

\subsection{Squeezed single-photon states}\label{sec:SSPS}

The squeezed single-photon states are defined as $S(\xi)|1\>$,  where $S(\xi)$ is the single-mode squeezing operator \eqref{smsop} and $|1\>$ is the one-photon Fock state \cite{kral}.
When the amount of squeezing is small, they are a good approximation of  the odd Schr{\"o}dinger cat states \cite{cat2,cat1,gran2007,telecat}, i.e.~of the superposition of two coherent states defined by
\begin{align*}
|\psi_{\rm odd\ cat}(\alpha)\>:=\frac {|\alpha\>-|-\alpha\>}{\sqrt{2\left(1-e^{-2|\alpha|^2}\right)}} \, .
\end{align*}

\begin{figure}[t]
\subfigure[]{
\includegraphics[height=3.3cm]{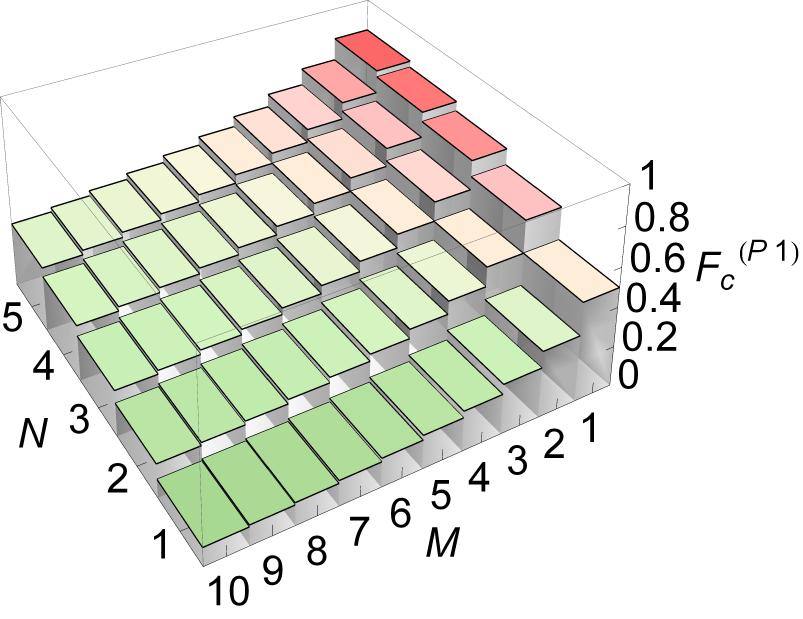}\label{p100}}\hspace*{.1cm}
\subfigure[]{
\includegraphics[height=3.3cm]{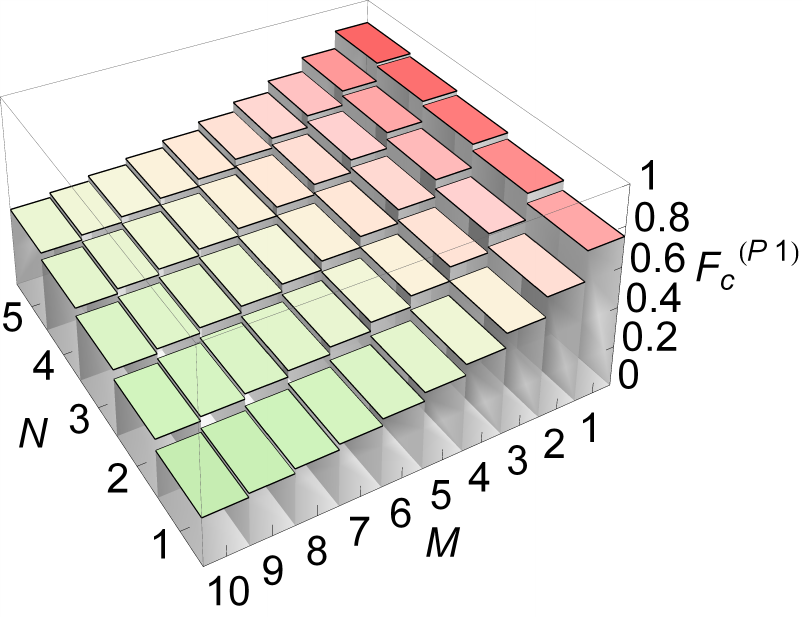}\label{p104}}\\
\subfigure[]{
\includegraphics[height=3.3cm]{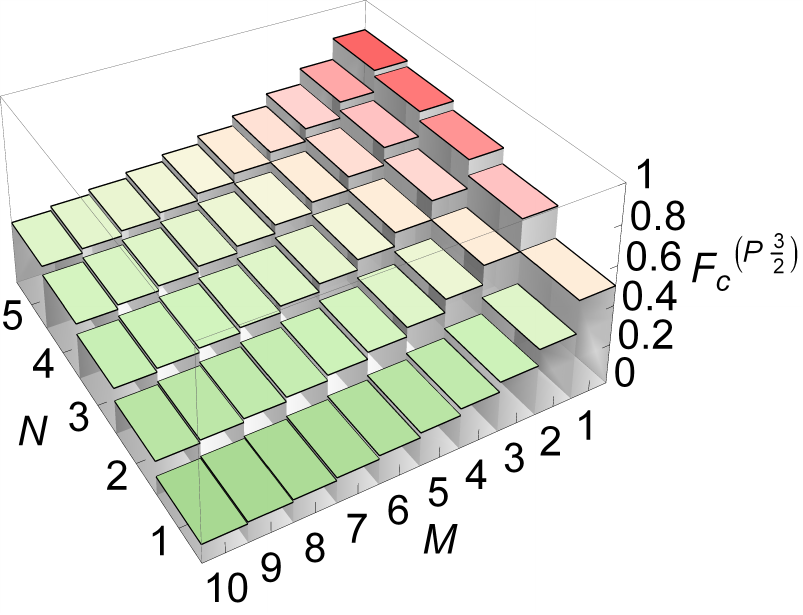}\label{p150}}\hspace*{.1cm}
\subfigure[]{
\includegraphics[height=3.3cm]{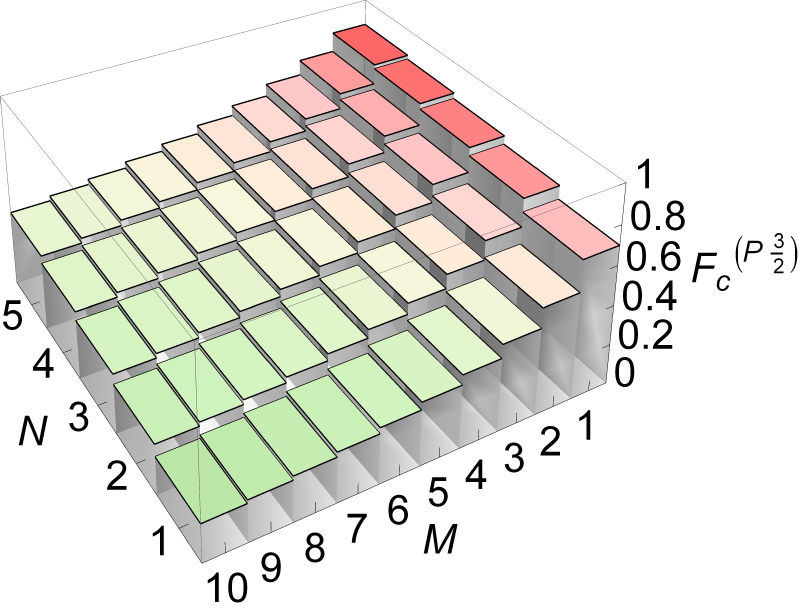}\label{p154}}\\
\subfigure[]{
\includegraphics[height=3.3cm]{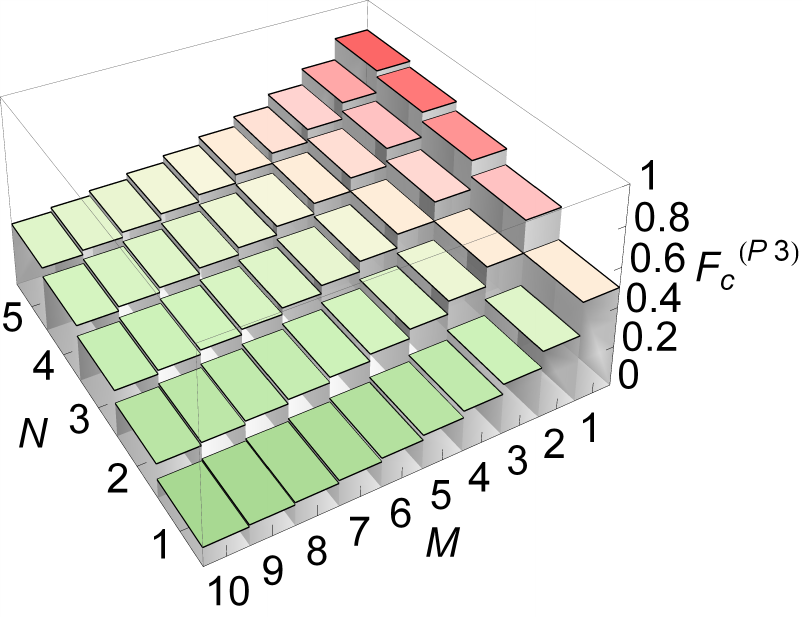}\label{p300}}\hspace*{.1cm}
\subfigure[]{
\includegraphics[height=3.3cm]{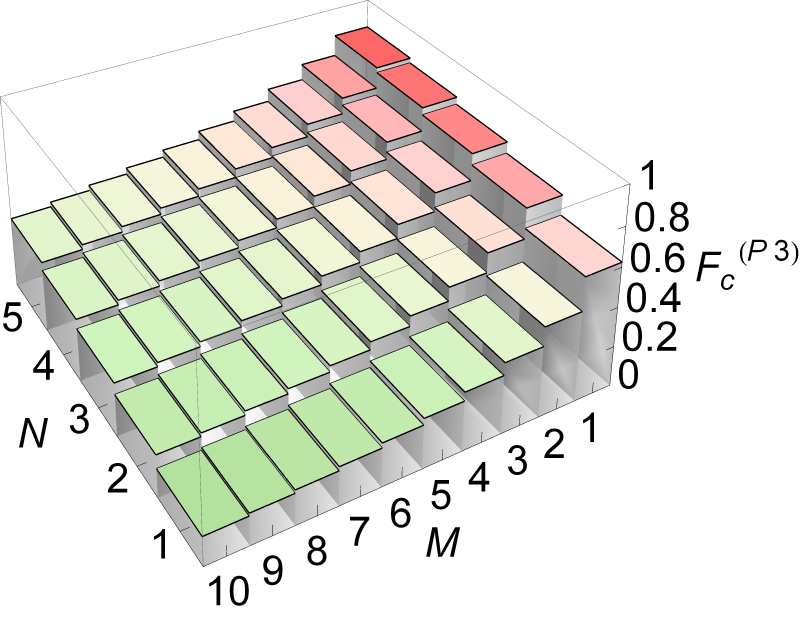}\label{p304}}\\
\caption{(Color online) Fidelity benchmark for the $N \rightarrow M$ transformation of Perelomov-$j$ squeezed states with (a)--(b) $j=1$, corresponding to (Gaussian) two-mode squeezed vacuum states, (c)--(d) $j=\frac{3}{2}$, corresponding to (non-Gaussian) squeezed single-photon states, and (e)--(f) $j=3$, corresponding to (non-Gaussian) states obtained by applying a two-mode squeezing operation to the Fock states $|2\>|0\>$.  The input states are distributed according to a prior distribution $p_\beta$ dependent on an inverse width parameter $\beta$ as explained in the main text. In the first column [panels (a), (c), (e)], we set $\beta=0$, corresponding to a uniform distribution. In the second column [panels (b), (d), (f)], we set $\beta=4$. The color legend for the CFT $F_{\rm c}^{({\rm P}\,j)}$ in the bar charts is: $0$ \protect\includegraphics[width=1.3cm]{fclegendf.pdf} $1$.
\label{fig:pjfell}}
\end{figure}

As we mentioned earlier,  the squeezed single-photon states are  the Perelomov-$j$  states with $j=3/2$.  Accordingly,  the benchmark for transforming $N$ copies of a squeezed single-photon state into $M$ copies is obtained by setting $j=k=3/2$ in Eq.  \eqref{eq:benchmarkperelomov}, which yields the CFT
\begin{align}\label{SSPS}
F_{\rm c}^{({\rm P}\frac32)}=\frac{3N+\beta}{3(N+M)+\beta}.
\end{align}
The benchmark  is plotted in Fig.~\ref{fig:pjfell}(c)--(d).  Note that when $\beta$ is large and the prior distribution  of the squeezing parameter is concentrated around $|\xi| = 0$, this benchmark can be approximately applied to the teleportation and cloning of odd Schr{\"o}dinger cat states.
The application of our benchmark leads to some rather strong consequences.  Indeed,  when the average photon number is small (here we choose $|\alpha|\le1$), an odd cat state $|\psi_{\rm odd\ cat}(\alpha)\>$ can be well approximated by the squeezed single-photon state $|\xi_*, 3/2\>$ where the optimal amount of squeezing is   $  \xi_*=s_*e^{2i\theta}$ with $s_* =\frac{1}{2}\ln\left(  |\alpha|^2/3+\sqrt{9+4|\alpha|^4}/3\right)$ and $\theta$ being the phase of $\alpha$  \cite{telecat}. Choosing $\beta$ to be large, one can guarantee that the squeezing degree is smaller than $|\xi_*  (\alpha = 1)|$ with high probability, so that the input states are likely to be close to odd cat states.  For example, one can consider the case of teleportation ($N=M=1$); solving the inequality
\begin{align*}
\int_0^{2\pi}\frac{\d\theta}{2\pi}\int_0^{s_*(1)}\frac{\beta\sinh s~\d s}{(\cosh s)^{\beta+1}}~\ge~0.99
\end{align*}
one gets $\beta\ge95.79$, which, plugged  into the benchmark \eqref{SSPS},   yields
\begin{align*}
F_{\rm c}^{(\rm odd\ cat)}\ge0.971.
\end{align*}
This lower bound shows that  MP protocols can achieve high fidelity in the teleportation of odd cat states with small photon numbers ($|\alpha|\le1$), due to the large amount of prior information available about the input.  In this regime, demonstrating a genuine quantum teleportation of odd cat states appears nearly impossible.  Of course, one can still consider the more feasible task given by teleportation of single-photon squeezed states  for broader prior distributions, allowing the input states to be different from cat states.

\subsection{Unbalanced two-mode squeezed number states}
In this Subsection we introduce the benchmark for another important family of continuous variable states, which are generated by applying the two-mode squeezing operator
\begin{align}\label{peum}
S^{(2)}(\xi)=\exp[-\xi^* a_1 a_2+\xi a_1^\dagger a_2^\dagger]
\end{align}
to   the two-mode Fock state $|m\>|0\>$, with $m$ an arbitrary integer.     For $m=0$, these are    the standard two-mode squeezed vacuum states, which are ubiquitous in continuous variable quantum information processing \cite{brareview,ourreview,pirandolareview}.
For $m  >0$, one has instead  a set of non-Gaussian unbalanced two-mode squeezed number states \cite{gerry}.
Like in the case of squeezed single-photon states, the  non-Gaussianity of these states is a precious resource for continuous variable quantum information processing \cite{nongaussian,gran2007,hybridrev}.  Consequently,   having a  benchmark for this class of states provides a useful criterion for the verification of quantum protocols empowered by non-Gaussianity \cite{nongaussparis}.

The unbalanced two-mode squeezed state of Eq.~(\ref{peum}) corresponds to the Perelomov-$(m+1)$ state
$$|\xi,  m+1\>=S^{(2)}(\xi)|m\>|0\>  \, ,$$
 where  the basis in  Eq.~\eqref{eq:perelomovstate} is $|\psi_n^{(m)}\>:=|n+m\>|n\>$.  Setting $k=j=m+1$ in Eq.~\eqref{eq:benchmarkperelomov}, we immediately get the CFT
\begin{align}
F_{\rm c}^{({\rm P}{\,m+1})}=\frac{2(m+1)N+\beta}{2(m+1)(N+M)+\beta}.
\end{align}
This is plotted in Fig.~\ref{fig:pjfell} for selected values of $m$ and $\beta$.

Notice in particular that the benchmark for Gaussian two-mode squeezed vacuum states ($m=0$) returns
\begin{align}
F_{\rm c}^{({\rm P}{\,1})}=\frac{2N+\beta}{2(N+M)+\beta}.
\end{align}
Such a CFT can be equivalently obtained from the benchmark for Gaussian single-mode squeezed vacuum states, Eq.~(\ref{benchSMSV}), by doubling $N$ and $M$. This is consistent with the fact that a two-mode squeezed vacuum state can be always generated from two single-mode squeezed vacuum states (equally squeezed in orthogonal quadratures) by means of a balanced beam splitter. A quantum memory for entangled two-mode squeezed light was demonstrated in \cite{fernnp}.  However, in that implementation the input squeezing degree was assumed to be known and therefore a different benchmark was used  to validate the experiment \cite{owari}.

\section{Deterministic vs probabilistic benchmarks}\label{SecSQ}
In the preceding sections we established the ultimate probabilistic benchmarks for a variety of protocols and input states. A natural question is whether there exist {\it deterministic} MP strategies which are able to achieve the benchmarks.
 In the following we address this question, considering separately the cases of uniform and non-uniform prior.

\subsection{Uniform prior:  optimality of the square-root measurement}

Let us assume that the input states $\{|\phi_g\>\}$  are chosen at random according to the normalized Haar measure.  This assumption is easily justified when the group $\grp G$ is compact. Nevertheless,  one can make sense of it also in  more general cases of non compact groups, by considering the uniform prior as the limit of a sequence of priors with increasing width. Alternatively, one can also interpret the (unnormalized) Haar measure as a weight that Victor assigns to the different input states when computing the figure of merit.

Whenever the input states are uniformly distributed (or weighted), one can prove that the the probabilistic benchmark can be reached by a deterministic strategy, which consists in performing  the \emph{square-root measurement} \cite{srm} and preparing the output state that corresponds to the measurement outcome.   In general, for a given set of states $\{\rho_x\}$ and a given set of probabilities $\{p_x\}$, the square-root measurement is the POVM
 $\{P_x\}_{x\in\mathsf X}$ defined by
\begin{align}\label{sqrtm}
P_x     =    \rho^{-\frac 12}\,    ( p_x \rho_x  )    \, \rho_x^{-\frac 12}   \, ,
\end{align}
where $\rho   =  \sum_{x}  p_x  \, \rho_x$ (note that this definition makes sense even if $\{p_x\}$ are  weights, provided that the sum is finite).
In the case of the uniformly-distributed GPCS   $\{  |\phi_g\>\}$, the square-root measurement is just the POVM with operators
\begin{align}\label{sqrtm}
P_g    =    d_\phi  \,   |\phi_g\>\<  \phi_g|   \qquad d_\phi   :  = \left(  \int  \d g \,    |\<  \phi  |  \phi_g\>|^2    \right)^{-1} \, .
\end{align}

In Appendix~\ref{app:optsquare} we prove that performing the square-root measurement and re-preparing the output state corresponding to the outcome is an optimal strategy.
 In other words, the ultimate CFT  $F_{\rm c}$  can be achieved with unit probability when the prior is uniform.
This feature holds also not only for MP protocols but also for the optimal quantum devices: in the case of uniform prior Alice and Bob can achieve the ultimate quantum fidelity $F_{\rm q}$ with probability 1 by applying a suitable quantum channel \cite{chiribellayang14}, which generalizes the universal cloning channel by Werner \cite{werner98} and the Gaussian cloners by Cerf, Ipe, Rottenberg, Iblisdir, and Lindblad \cite{CerfIpe2000,CerfIblisdir2000,Lindblad2000}.

\subsection{Non-uniform prior: a counterexample}\label{SubSecSQ}

The result of the previous paragraph might suggest that one can \emph{always} attain our benchmarks using  a deterministic protocol based on the square-root measurement.
  A further clue that goes in that direction comes from  the  benchmarks for coherent states  of the harmonic oscillator \cite{hammerer,namikiprl,giulionew}, which can all be achieved by the square-root measurement.
 Nevertheless, these results cannot be generalized to arbitrary GPCS:  here we show a counterexample where  the square-root measurement does not achieve the  probabilistic benchmark.
It remains however an open question whether more complicate deterministic strategies can achieve  our probabilistic benchmarks in the general case.

Suppose that Alice wants to teleport a single-qubit state to Bob, chosen at random on the Bloch sphere according to the non-uniform prior  $p_\beta  (\d g)    =   d_\beta  |   \<  \phi |  \phi_g \>|^{2\beta}  \d g$ of Eq.~(\ref{qubitprior}) and Fig.~\ref{fig:qubitpriorall}.
Consider the  MP protocol  that consists in measuring the square-root measurement with POVM operators
\begin{align}\label{sqrtPOVM}
P_\eta(g):=p_\eta(g)~\rho_\eta^{-\frac{1}{2}} \, (|\psi_g\>\!\<\psi_g|)^{\otimes N} \, \rho_\eta^{-\frac{1}{2}},
\end{align}
where
$$\rho_\eta=\int p_\eta(\d g) \,    \left(  |\psi_{g}\>\!\<\psi_{g}|\right)^{\otimes N}  \, .$$
Strictly speaking, the square-root measurement associated to the input ensemble would have $\eta  \equiv \beta$, but here
 we grant Alice and Bob the additional freedom to  optimize over   $\eta $, possibly obtaining a better fidelity.

\begin{figure}
\centering
\includegraphics[width=8.5cm]{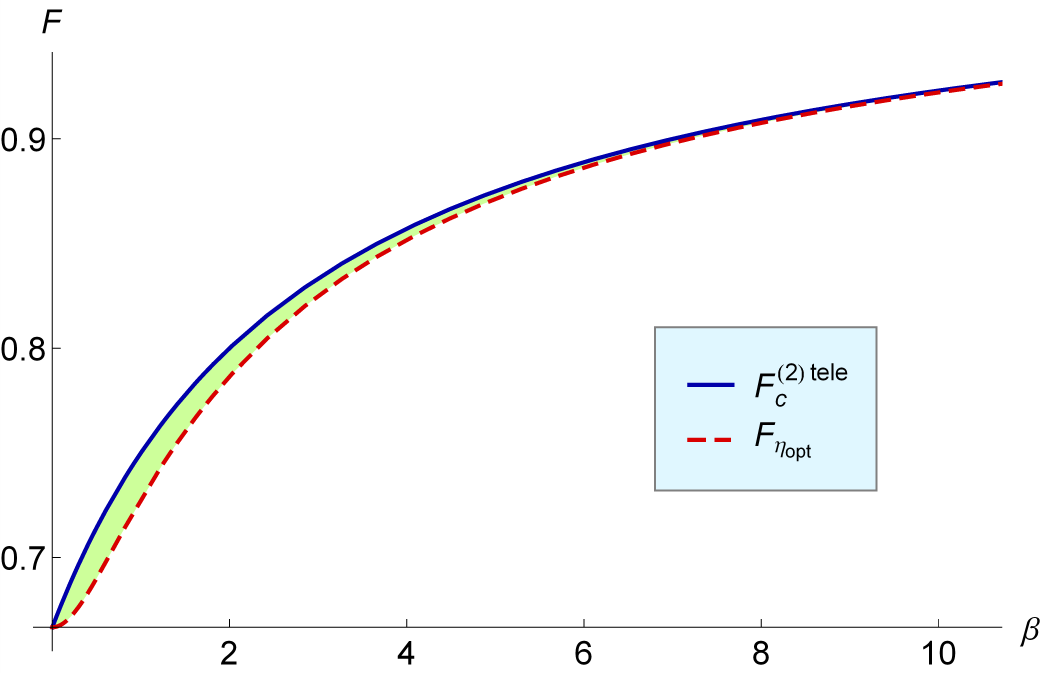}
\caption{(Color online) The fidelity $F_{\eta_{\rm opt}}(\beta):=\max_\eta F(\beta,\eta)$ for the optimal deterministic square-root measurement (red dashed line) and the ultimate probabilistic benchmark $F_{\rm c}^{(2){\rm tele}}(\beta)$ for one-qubit teleportation (blue solid line) as functions of the inverse width $\beta$ of the prior distribution. The (green) filling between the two lines represents the gap  between the benchmark and the performance of the optimal square-root measurement.}\label{fig:squareroot}
\end{figure}

By performing the square-root measurement with parameter $\eta$ and re-preparing the qubit state corresponding to the outcome, Alice and Bob get the teleportation fidelity
\begin{align}\label{sqrtfidelity}
F(\beta,\eta)&=\int \d g~\int \d\hat{g}~p_\beta(g)~p_\eta(\hat{g})~\left|\<\psi_g|\tau_\eta^{-\frac{1}{2}}|\psi_{\hat{g}}\>\right|^2~\left|\<\psi_g|\psi_{\hat{g}}\>\right|^{2}.
\end{align}
For fixed inverse width $\beta$, we can now find the value of $\eta$ that maximizes the teleportation fidelity.  The optimization can be done analytically (see Appendix \ref{sec:proofsquareroot}) and yields a   fidelity $F_{\eta_{\rm opt}}(\beta)$ that is strictly smaller than the benchmark   $F_{\rm c}^{(2){\rm tele}}(\beta)=\frac{\beta+2}{\beta+3}$ from  Eq. (\ref{qubittele}).
In Fig.~\ref{fig:squareroot}, the fidelity for the optimal square-root measurement  $F_{\eta_{\rm opt}}(\beta)$ and the benchmark $F_{\rm c}^{(2){\rm tele}}$ are plotted as functions of $\beta$. Clearly,  the square-root measurement is not optimal for every finite value   $\beta  \in  (0,\infty)$.
The explicit expression of the gap $\Delta(\beta)$ between the two fidelities  is reported in Appendix \ref{sec:proofsquareroot}.
Interestingly, for large $\beta$ the gap vanishes polynomially with $\beta$, precisely $\Delta(\beta) \approx O\left(\beta^{-3}\right)$. Therefore we conclude that the square-root measurement is asymptotically optimal, when the prior knowledge about the input state becomes more and more peaked.

\section{Quantum benchmarks vs optimal quantum strategies: a task with no quantum advantage}\label{SecTrans}

The main application of quantum benchmarks is the certification of quantum advantages in realistic experimental settings.  However, not every task displays a quantum advantage:  for some transformations, the maximum fidelity achieved by arbitrary quantum strategies is equal to the quantum benchmark associated to the best MP procedure.

An example of this situation is the following:
Suppose that we want to transform $N$ copies of the GPCS  $  |\phi_g\>$, distributed with prior probability $  p_{\gamma}  (g)    =d_\gamma \,   |  \< \phi_{\gamma} |  \phi_{\gamma,g}\>|^2  $, into $M$ copies of the state $  |\psi^*_g\>$, the complex conjugate of another GPCS state $  |\psi_g\>$.
As in the rest of the paper, we assume that the states $ |\psi_g\>^{\otimes M}$,    $ |\phi_g\>^{\otimes N}$, and $ |\phi_{\gamma,g}\>$ are mutually coherent, i.e.~that their product is a GPCS.
In this setting, we can prove  that there is no quantum advantage:  MP protocols are optimal among all possible quantum strategies.  The proof is presented in Appendix   \ref{sec:cc}, where we  show the equality  $F_{\rm q}   =  F_{\rm c}$ between the ultimate quantum fidelity in Eq.~(\ref{qfid}) and the benchmark of Eq.~(\ref{bench}).

The simplest example of this situation is illustrated by  the optimal approximation of the universal {\tt NOT} gate for qubits \cite{BuzekTranspose,demartiniexp}, which is unitarily equivalent to the optimal complex conjugation (i.e., optimal transposition).  In this case, the fact that a MP protocol is the best among all possible quantum strategies was exploited  in Ref.~\cite{TransposePRL} to design an experimental implementation of the optimal transposition of photonic qubit states.
For higher dimensional systems and for coherent states of the harmonic oscillator,  the optimal transposition map was derived in Ref.~\cite{buscemi2003}.  Very recently, the observation  that this map can be achieved  by a MP protocol has been used for an experimental demonstration for qutrit states encoded into the path and polarization degrees of freedom of a single photon \cite{KalevBae2013}.  Optimal MP protocols for the transposition of arbitrary pure states in finite dimensions have also been proposed in   \cite{KalevBae2013}.
Our result extends the optimality of MP protocols to a large variety of new scenarios, including the transposition of various Gaussian and non-Gaussian pure states, distributed according to a generally  non-uniform prior.


\section{Discussion and conclusions}\label{SecCC}
Very recent experiments have reached a remarkable degree of control in the preparation and manipulation of hybrid states of light and matter \cite{gran2007,furunature2013,hybridJeong,hybridMorin}. Hybrid optical technologies \cite{hybridrev} combine the best of both worlds, discrete variable and continuous variable systems, in order to efficiently realize elementary building blocks for distributed quantum communication \cite{QIkimble}. For instance, in a recent experiment a single-photon state was teleported unconditionally, i.e., deterministically, using a continuous variable setup \cite{furunature2013}. Complementary to such an approach, a proposal to realize high-fidelity teleportation of continuous variable states by multiple single-photon teleportation channels was also recently put forward \cite{anderralph}.
One can easily imagine extensions of such protocols from teleportation to the more general instance of asymmetric telecloning \cite{telecloning}, where remote parties receive approximate copies of input unknown states, possibly with nonuniform fidelities. So far, telecloning has been successfully implemented for Gaussian (coherent) inputs exploiting Gaussian shared entangled states \cite{exptelecloning}.

However, to the best of our knowledge, a completely satisfactory teleportation or telecloning experiment demonstrating the use of entangled resources to surpass the best classical performance in the transmission of an {\it ensemble} of input states has not been achieved to date. Such an experiment would require sampling a high number of input states from a prior distribution known   to senders and receivers, accomplishing the transmission for each  input, and eventually calculating the experimental average fidelity over the ensemble. This should then be compared to, and shown to surpass, the ultimate  benchmark corresponding to the given input ensemble.

The present work offers a powerful and general machinery to derive such benchmarks for many instances of relevant input ensembles, with adjustable prior distributions which can be tailored to reliably mimic the experimental facilities. One of the novelties and strenghts of our approach to quantum benchmarking, is that it encompasses under a unifying mathematical framework both discrete and continuous variable states, and both Gaussian and non-Gaussian ones in the latter case. The obtained benchmarks are therefore ideally and naturally suited to be used as tests to validate current and future implementations of hybrid quantum communication.

We hope this paper may stimulate further efforts by the experimental community, and further interaction with the theoretical one, in order to reach new heights in the successes of quantum optical technologies and demonstrate performances even more markedly above what classically possible. In particular, new (possibly probabilistic) protocols and new tools are required to beat the benchmarks for the transmission of squeezed vacuum or squeezed Fock states (or, similarly, Schr\"odinger cat-like states) using affordable degrees of shared entangled resources, as discussed e.g.~in \cite{kogias14} and in the present paper.

Finally, despite the large variety of relevant cases in which this work succeeds to deliver analytical benchmarks, let us briefly comments on some limitations of our approach which call for further investigation. One of the main limitations is in the fact that our input states are always assumed to be pure. A more faithful analysis of experiments might require the derivation of benchmarks for ensembles of initially mixed states; while this may not be possible in closed form for all the classes of states analyzed here, one can always resort to numerical methods, e.g.~semidefinite programming as in \cite{mariona}. From this point of view, Eq. (\ref{bench}) represents a convenient starting point.   More generally, the application of our group-theoretic methods to the assessment of the ultimate quantum (rather than classical) bounds in protocols such as remote state preparation, optimal quantum cloning, amplification and attenuation of generalized coherent states, and optical imaging technologies, will be analyzed in forthcoming works.

\acknowledgments{
This work was supported   by the National Basic Research Program of China
(973) 2011CBA00300 (2011CBA00301), by the National
Natural Science Foundation of China (Grants 11350110207,
61033001, 61061130540), by the 1000 Youth Fellowship
Program of China, by the
Tsinghua--Nottingham Teaching and Research Fund 2013, by the Foundational Questions Institute (Grants FQXi-RFP3-1325 and FQXi-RFP3-1317), and by the Brazilian CAPES (Pesquisador
Visitante Especial-Grant~No. 108/2012).
We thank I. Kogias, S. Ragy, and U. Andersen for fruitful discussions. GA acknowledges the hospitality of Tsinghua University.  GC acknowledges the hospitality of the Simons Institute for the Theory of Computing and of Perimeter Institute for Theoretical Physics.
Research at Perimeter Institute for Theoretical Physics is supported in part by the Government of Canada through NSERC and by the Province of Ontario through MRI.}

\appendix
\begin{widetext}
\section{Proofs}

\subsection{Proof of Theorem~\ref{theo:bench}}
Since we have already exhibited a probabilistic MP protocol that achieves the fidelity in Eq.~(\ref{probCFT}), here we only need to show that the r.h.s. of Eq.~(\ref{probCFT}) is an upper bound for arbitrary MP protocols.  To this purpose, we start from Eq.~(\ref{bench}), which now reads
\begin{align}\label{eq:fidelity}
F_{c}(\gamma)  =   \left\|  \left(  I_{\rm out}  \otimes   \rho_\gamma^{-\frac 12}\right)     \Omega_\gamma   \left(  I_{\rm out}  \otimes   \rho_\gamma^{-\frac 12}\right)   \right\|_\times
\end{align}
where $\rho_\gamma$ is the averaged input state, given by
\begin{align*}
\rho_\gamma &:=\int p_\gamma(\d g)~\left(|\phi_g\>\!\<\phi_g|\right)^{\otimes N}
\end{align*}
and $\Omega_\gamma$ is the average input-output state, given by
\begin{align*}
\Omega_\gamma &: =    \int  \d g  \,    p_\gamma (g)  \,     (|\psi_g\>\<  \psi_g|)^{\otimes M}  \otimes  (|\phi_g\>\<  \phi_g|)^{\otimes N}
\end{align*}
is the averaged  input-output state.  Now, for every operator $A$  we have the inequality $   \|  A  \|_\times \le \| A\|_\infty$, where $\|A  \|_\infty =   \sup_{  \|  |\Psi\> \|  = 1}   \< \Psi |A |\Psi\>$.
Hence, we have the bound
\begin{align*}
F_{\rm c}(\gamma)  &\leqslant \left\| \left(  I_{\rm out}  \otimes   \rho_\gamma^{-\frac 12}\right)     \Omega_\gamma   \left(  I_{\rm out}  \otimes   \rho_\gamma^{-\frac 12}\right) \right\|_\infty\\
&=\min \left\{ \lambda \ge 0 ~ | ~     \left( I_{\rm out}  \otimes   \rho_\gamma^{-\frac 12}\right)     \Omega_\gamma   \left(  I_{\rm out}  \otimes   \rho_\gamma^{-\frac 12}\right)       \le \lambda   \, I_{\rm out}  \otimes  I_{\rm in}  \right\} \\
&=\min\left\{\lambda\geqslant0~|~  \Omega_\gamma  \le \lambda(I_{\rm out}\otimes\rho_\gamma) \right\}\\
&  =  : \lambda_{\min}
\end{align*}
On the other hand, $\rho_\gamma$ and $\Omega_\gamma$ can be expressed  as
\begin{align*}
\rho_\gamma &=d_\gamma \Tr_{\spc H_\gamma}\Bigg[\left(\int \d g~(|\phi_g\>\!\<\phi_g|)^{\otimes N}\otimes|\phi_{\gamma,g}\>\!\<\phi_{\gamma,g}|\right)\left(I_{\rm in}\otimes|\phi_\gamma\>\!\<\phi_\gamma|\right)\Bigg]\\
\Omega_\gamma &=d_\gamma \Tr_{\spc H_\gamma}\Bigg[\left(\int \d g~(|\psi_g\>\!\<\psi_g|)^{\otimes M}\otimes(|\phi_g\>\!\<\phi_g|)^{\otimes N}\otimes|\phi_{\gamma,g}\>\!\<\phi_{\gamma,g}|\right)  \left(I_{\rm out} \otimes I_{\rm in}  \otimes|\phi_\gamma\>\!\<\phi_\gamma|\right)\Bigg] \, .
\end{align*}
Since by hypothesis the  states  $|\phi_g\>^{\otimes N},   |\psi_g\>^{\otimes M}$ and $|\phi_{\gamma,g}\>$ are mutually coherent, the Schur's lemma implies that the integrals in the r.h.s. of the above equations are  proportional to  projectors on irreducible subspaces. Specifically, we have
\begin{align}\label{aaa}
\int \d g~(|\phi_g\>\!\<\phi_g|)^{\otimes N}\otimes|\phi_{\gamma,g}\>\!\<\phi_{\gamma,g}|   & =  \frac{P}{d_P}   \qquad \qquad d_P:     = \left( \int \d g  \,     |\< \phi|\phi_g\> |^{2 N}   |\<  \phi_\gamma|\phi_{\gamma,g}\>|^2\right)^{-1} \\
\label{bbb}
\int \d g~(|\psi_g\>\!\<\psi_g|)^{\otimes M}\otimes(|\phi_g\>\!\<\phi_g|)^{\otimes N}\otimes|\phi_{\gamma,g}\>\!\<\phi_{\gamma,g}|   &=  \frac{Q}{d_Q}  \qquad \qquad d_Q:     =  \left(\int \d g  \,       |\< \psi|\psi_g\> |^{2 M}        |\< \phi|\phi_g\> |^{2 N}   |\<  \phi_\gamma|\phi_{\gamma,g}\>|^2\right)^{-1}  \, ,
\end{align}
where $P$ and $Q$ are the projectors on the irreducible subspaces  $\spc H_P$ and $\spc H_Q$ spanned by the vectors $\{|\phi_g\>^{\otimes N}  |\phi_{\gamma,g}\>\}$ and  $\{|\psi_g\>^{\otimes M}  \>|\phi_g\>^{\otimes N}  |\phi_{\gamma,g}\>\}$, respectively.
Since $\spc H_Q$ is a subspace of $\spc H^{\otimes M}  \otimes \spc H_P$, we have $Q  \le (I^{\otimes M}  \otimes    P)$.  Hence, we have the bound
\begin{align*}
\Omega_\gamma    &=d_\gamma\Tr_{\spc H_\gamma}\left[\frac{Q}{d_Q}(I_{\rm out} \otimes I_{\rm in}  \otimes|\phi_\gamma\>\!\<\phi_\gamma|)\right]   \\
& \le   d_\gamma\Tr_{\spc H_\gamma}\left[\frac{ I_{\rm out}  \otimes P}{d_Q}(I_{\rm out} \otimes I_{\rm in}  \otimes|\phi_\gamma\>\!\<\phi_\gamma|)\right] \\
&=\frac{d_P}{d_Q}    \,      I_{\rm out} \otimes \left\{    d_\gamma\Tr_{\spc H_\gamma}\left[\frac{P }{d_P}(I_{\rm in}\otimes |\phi_\gamma\>\!\<\phi_\gamma|)\right]\right\}\\
& =  \frac{d_P}{d_Q}    \,    I_{\rm out} \otimes \rho_\gamma  \, .
\end{align*}
In conclusion, we have proven the inequality
\begin{align}\label{uppe}
\lambda_{\min} \le
   \frac{d_P}{d_Q}  =  \frac{\int \d g  \,       |\< \psi|\psi_g\> |^{2 M}        |\< \phi|\phi_g\> |^{2 N}   |\<  \phi_\gamma|\phi_{\gamma,g}\>|^2}{\int \d g  \,           |\< \phi|\phi_g\> |^{2 N}   |\<  \phi_\gamma|\phi_{\gamma,g}\>|^2} \, ,
\end{align}
having used Eqs. (\ref{aaa}) and (\ref{bbb}) in the second equality.

Combined with the inequality $ F_{\rm c}(\gamma) \le \lambda_{\min}$, Eq. (\ref{uppe})  gives the desired upper bound once.
Since the r.h.s. of the inequality is achievable by a probabilistic MP protocol (see the main text),   we proved the following equality
\begin{align}\label{equivalence}
F_{\rm c}     (\gamma)  =   \left\|  \left(  I_{\rm out}  \otimes   \rho_\gamma^{-\frac 12}\right)     \Omega_\gamma   \left(  I_{\rm out}  \otimes   \rho_\gamma^{-\frac 12}\right)   \right\|_\times  \equiv     \left\|  \left(  I_{\rm out}  \otimes   \rho_\gamma^{-\frac 12}\right)     \Omega_\gamma   \left(  I_{\rm out}  \otimes   \rho_\gamma^{-\frac 12}\right)   \right\|_\infty  \, .
\end{align}
\hfill \qed

\subsection{Proof of the benchmark  in Eq.~(\ref{eq:benchmarkqudit}) for general $d$  and general $\beta$.}\label{sec:proofqudit}
Eq.~(\ref{eq:benchmarkqudit})  is the specialization of Eq.~(\ref{probCFT}) to pure qudit states.  Since we know that the fidelity in   Eq.~(\ref{probCFT}) is achievable,  here we only need to show that no MP protocol can achieve a higher fidelity.  To this purpose, we use the upper bound
\begin{align}\label{aaa}  F_{\rm c}  \le \left  \|   A \right\|_\infty   \qquad \qquad A  :  =    \left(I_{\rm out}  \otimes   \rho^{-\frac 12}\right) \Omega     \left(I_{\rm out}  \otimes   \rho^{-\frac 12}\right)   \, ,
\end{align}
which follows immediately from  the general expression for the CFT given in Eq.~(\ref{bench}) and from the inequality $\|  A\|_\times \le \|  A\|_\infty$.
In the following  we compute directly the operator norm, by working out the expressions for the  operators   $\rho$ and  $\Omega$.   Since these operators depend on the inverse width $\beta$, we  will denote them by   $\rho_\beta$ and  $\Omega_\beta$, namely
\begin{align}
\nonumber \rho_\beta  &:  =  \int\d \psi ~p_\beta(\psi)~\left(  |\psi\>\!\<\psi|\right)^{\otimes N} \\
\nonumber \Omega_\beta  &:  =  \int\d \psi ~p_\beta(\psi)~\left(  |\psi\>\!\<\psi|\right)^{\otimes {(M+N)}}\\
\label{ccc}
A_\beta  &  :  =    \left(I_{\rm out}  \otimes   \rho_\beta^{-\frac 12}\right) \Omega_\beta     \left(I_{\rm out}  \otimes   \rho_\beta^{-\frac 12}\right)  \, .
\end{align}

Let us start from the computation of $\rho_\beta$ (the computation of $\Omega_\beta$ is identical and the result can be obtained by replacing $N$ by $M+N$).
Using the Hurwitz parametrization, the state $|\psi\>^{\otimes N}$ can be expressed as
\begin{align}\label{bbb}
|\psi\>^{\otimes N}=\sum_{\st{n} \in \grp{P}_{N,d}  }\sqrt{N\choose\st{n}}\left(\prod_{j=0}^{d-1} c_j   \right)|N,\st{n}\>.
\end{align}
where the summation is over the set $\grp P_{N,d}$ consisting of all partitions of $N$ into $d$ non-negative integers,   ${N\choose \st{n}}$ is the multinomial
$${N\choose \st{n}}:=\frac{\Gamma(N+1)}{\prod_{j=0}^{d-1}\,   \Gamma(n_j+1)} \,   ,$$   the coefficients   $\{c_j\}$  are given by
\begin{align*}
c_0   &  := e^{i\theta_0}   \cos \theta_0  \\
c_j&:= e^{i\theta_j}\cos  \theta_j  \left(  \prod_{k=0}^{j-1}   \sin\theta_{k} \right)   \qquad j  \in  \{1,\dots, d-2\}\\
 c_{d-1}   & :  =    \prod_{k=0}^{d-2}   \sin\theta_{k}  \, .
\end{align*}
and $ |N,  \st n\>$ is the unit vector  obtained by projecting the vector $ |0\>^{n_0} |1\>^{n_1}  \cdots  |d-1\>^{n_{d-1}}$ on  the symmetric subspace.
Using Eqs. (\ref{bbb}), (\ref{quditprior}) and (\ref{sudhaar}),  we then obtain
\begin{align*}
\rho_\beta&=        d_\beta \,  (d-1)! \, 2^{d-1}     \sum_{\st{n}  \in \grp{P}_{N,d}}{N\choose \st{n}}\left[\int_0^{\pi/2}\d\theta_{0}(\cos\theta_{0})^{2\beta+1}(\sin\theta_{0})^{2(d-1)-1} \, \prod_{j=1}^{d-2}\int_0^{\pi/2}\d\theta_j\cos\theta_{j}(\sin\theta_{j})^{2(d-j-1)-1}\right]~\left|\prod_{j=0}^{d-1} c_j^{n_j}\right|^2 \, |N,\st{n}\>\!\<N,\st{n}|\\
&=   d_\beta \,  (d-1)! \, 2^{d-1}    \sum_{\st{n} \in \grp{P}_{N,d}}{N\choose \st{n}}\left[\int_0^{\pi/2}\d\theta_{0}(\cos\theta_{0})^{2(\beta+1 +n_0)-1}(\sin\theta_{0})^{2\left(d-1 +\sum_{k=1}^{d-1} n_{k}\right)-1}\right]  \,  \times  \\
   &  \qquad \qquad \qquad \quad \quad~~~~~~~~~\times \left[
\prod_{j=1}^{d-2}\int_0^{\pi/2}\d\theta_j(\cos\theta_{j})^{2(n_j+1)-1}  (\sin\theta_{j})^{2\left(d-j -1 +  \sum_{k=j+1}^{d-1}n_k\right)-1}\right]  
  \, |N,\st{n}\>\!\<N,\st{n}|  \, .
 \end{align*}
The integrals can be computed using the relation
 \begin{align*}
 \int_0^{\pi/2}\d x  \,  (\cos x )^{2p-1}(\sin x )^{2q-1}   =  \frac{  \Gamma (t)  \Gamma(q)}{  2  \Gamma(p+q)}  \, ,
 \end{align*}
which yields
  \begin{align}
\nonumber \rho_\beta&=   d_\beta \,  (d-1)! \, 2^{d-1}  \sum_{\st{n} \in \grp{P}_{N,d}}{N\choose \st{n}}  \, \left[  \frac{  \Gamma( \beta  +1+  n_0 )  \Gamma( d  -1 + \sum_{k=1}^{d-1}n_{k}  )}{2\Gamma( \beta  +d+   N)}   \right]\left[
\prod_{j=1}^{d-2}
  \frac{\Gamma( n_j  +1)  \Gamma( d-j  -1 +  \sum_{k=j+1}^{d-1}  n_k )}{ 2  \Gamma( d- j  +  \sum_{k=j}^{d-1}  n_k)}
  \right]  \, |N,\st{n}\>\!\<N,\st{n}|\\
\nonumber  &=      d_\beta \,  (d-1)!     \sum_{\st{n} \in \grp{P}_{N,d}}{N\choose \st{n}}   \, \frac{\Gamma(\beta+n_0+1)   \, \prod_{j=1}^{d-1}\Gamma(n_j+1)}{\Gamma(\beta+d+  N)}  \, |N,\st{n}\>\!\<N,\st{n}|\\
\label{eee}
&=d_\beta    \,  \sum_{\st{n} \in \grp{P}_{N,d}}   \frac{{N\choose \st{n}}}{   {{N+\beta+d-1}\choose{d-1}}{{N+\beta}\choose{\st{n}+\bs{\beta}}}  } \, |N,\st{n}\>\!\<N,\st{n}|   \qquad \qquad \bs{\beta}:=(\beta,0,\dots,0)  \in\R^{d} \, .
\end{align}
Replacing $N$ by $M+N$, we obtain the expression for the operator $\Omega_\beta$
\begin{align}\label{ddd}
\Omega_\beta&=d_\beta\sum_{\st{t} \in \grp{P}_{M+N,d}}  \frac{{M+N \choose \st{t}}}  {{{M+N+\beta+d-1}\choose{d-1}}{{M+N+\beta}\choose{\st{t}+\bm{\beta}}} }  \, |M+N,\st{t}\>\!\<M+N,\st{t}|  \, .
\end{align}

In order to compute the operator norm in Eq.~(\ref{aaa}), we now diagonalize the operator $A_\beta$.
To this purpose, we define the vectors
\begin{align*}
|\Phi_{M+N, \st t}  \>  :    =    \left( I_{\rm out}  \otimes \rho_\beta^{- \frac 12}\right)   |M+N,   \st t\> \, ,
\end{align*}
so that one has
\begin{align}\label{fff}
A_{\beta}   = d_\beta\sum_{\st{t} \in \grp{P}_{M+N,d}}  \frac{{M+N \choose \st{t}}}{  {{M+N+\beta+d-1}\choose{d-1}}{{M+N+\beta}\choose{\st{t}+\bm{\beta}}}  } \, |\Phi_{M+N,\st{t}}\>\!\<\Phi_{M+N,\st{t}}| \, ,
\end{align}
as follows from Eqs. (\ref{ccc}) and (\ref{ddd}).
Note that the vectors $  |\Phi_{M+N,\st t}\>$   are mutually orthogonal.  Indeed, using the relation
\begin{align*}
|M+N, \st t  \>     =  \sum_{
\begin{array}{c}
\st m   \in  \grp  P_{M,d}   \, ,\st n\in  \grp P_{N,d} \\
\st m+  \st n  =  \st t
\end{array}}  \sqrt{ \frac{  {M \choose \st{m}} {N \choose \st{n}}  }{{M+N \choose \st{t}}} }    \,   |M,\st m\>  |N,\st n\>
\end{align*}
we get
\begin{align*}
\left \<  \Phi_{M+N,\st t}  |  \Phi_{M+N,\st t'}   \right\> &  =   \<  M+N,   \st t|     \left( I_{\rm out}  \otimes \rho_\beta^{- 1}\right)   |M+N,   \st t'\>\\
&  =  \sum_{\begin{array}{c}\st m+  \st n    =  \st t\\
\st m'  + \st  n'  =   \st t'
\end{array}}      \sqrt{ \frac{  {M \choose \st{m}} {N \choose \st{n}}  }{{M+N \choose \st{t}}} }         \sqrt{ \frac{  {M \choose \st{m'}} {N \choose \st{n'}}  }{{M+N \choose \st{t'}}} }   \,    \<  M, \st m  | M,  \st m'\>  \,    \<  N , \st  n | \,  \rho_\beta^{-1}\,  |N,\st n'\>  \\
&  =  \sum_{\begin{array}{c}\st m+  \st n    =  \st t\\
\st m'  + \st  n'  =   \st t'
\end{array}}      \sqrt{ \frac{  {M \choose \st{m}} {N \choose \st{n}}  }{{M+N \choose \st{t}}} }         \sqrt{ \frac{  {M \choose \st{m'}} {N \choose \st{n'}}  }{{M+N \choose \st{t'}}} }   \,    \delta_{\st m,\st m'} \,         \left[   d_\beta^{-1}      \frac{{{N+\beta+d-1}\choose{d-1}}{{N+\beta}\choose{\st{n}+\bs{\beta}}}}{{N\choose \st{n}}}  \,  \delta_{\st n,\st n'}   \right] \\
&  = \delta_{\st t,\st t'}  \,    \left[    \sum_{\st m+  \st n    =  \st t  }        \frac{  {M \choose \st{m}}   {{N+\beta}\choose{\st{n}+\bs{\beta}}}   }{{M+N \choose \st{t}}}   \right]       \,   d_\beta^{-1}          {{N+\beta+d-1}\choose{d-1}}   \\
&  = \delta_{\st t,  \st t'}\,       \frac{  {M  +  N  +  \beta \choose \st{t}  +  \bs \beta} }{{M+N \choose \st{t}}}          \, d_\beta^{-1}      {{N+\beta+d-1}\choose{d-1}} \, .
 \end{align*}
Here we used Eq.~(\ref{eee}) in the third equality and the Chu-Vandermonde identity for multinomials \cite{multinomial}.
 in the fifth equality.

Finally, defining the normalized vectors
$$  |  \overline{\Phi}_{  M+N,\st t}\>  :  =  \frac{  |  {\Phi}_{  M+N,\st t}\>  }{\sqrt{\<  \Phi_{M+N,\st t}  |  {\Phi}_{  M+N,\st t}\> }}$$  and using Eq.~(\ref{fff}) we obtain the desired diagonalization
\begin{align*}
A_{\beta}  =  \frac{{{N+\beta+d-1}\choose{d-1}}}{{{M+N+\beta+d-1}\choose{d-1}}}  \sum_{\st{t} \in \grp{P}_{M+N,d}}|\overline{\Phi}_{M+N,\st{t}}\>\!\<\overline{\Phi}_{M+N,\st{t}}| \, .
\end{align*}
Note that the operator $A_\beta$ is proportional to a projector and the proportionality constant is the operator norm
$$ \left\|A_\beta  \right\|_\infty     =    \frac{{{N+\beta+d-1}\choose{d-1}}}{{{M+N+\beta+d-1}\choose{d-1}}}    \, . $$

By Eq.~(\ref{aaa}),    we obtain  the upper bound $F_{\rm c}  \le {{N+\beta+d-1}\choose{d-1}}/ {{M+N+\beta+d-1}\choose{d-1}}$.  The upper bound coincides with the achievable  value given in Eq.~(\ref{eq:benchmarkqudit}), thus proving the optimality of our benchmark.

\subsection{Proof of the benchmark for multimode squeezed states Eq.~(\ref{eq:benchmarkperelomov}).}\label{SecProofMultimode}
In this section we prove the benchmark Eq.~(\ref{eq:benchmarkperelomov}) following a similar route as we did in Appendix~\ref{sec:proofqudit}. We will proved an upper bound for the MP protocols using Eq.~(\ref{aaa}), and then show that it is achievable.

In order to compute the operator norm in Eq.~(\ref{aaa}), we need to first determine $\rho_\beta$ and $\Omega_\beta$ (labeled by the inverse width parameter $\beta$ of the prior). A tensor product of $M$ identical copies of Perelomov-$k$ states and $N$ identical copies of Perelomov-$j$ states can be represented as
\begin{align}\label{Ncopy}
|\xi,k\>^{\otimes M}|\xi , j \>^{\otimes N}=\frac{1}{(\cosh s)^{Mk+Nj}}\sum_{n=0}^{\infty}(e^{i\theta}\tanh s)^{n}\sqrt{{kM+jN+n-1\choose n}}\left|\Psi_{M,N,n}^{(k,j)}\right\>.
\end{align}
Here $\left\{\left|\Psi_{M,N,n}^{(k,j)}\right\>\right\}_{n}$ is a set of vectors defined as
\begin{align*}
\left|\Psi_{M,N,n}^{(k,j)}\right\>=\sqrt{{kM+jN+n-1\choose n}^{-1}}\sum_{\{n_i\}\in\grp P_{n,M+N}}\left(\bigotimes_{\alpha=1}^{M}\sqrt{{n_\alpha+k-1\choose n_\alpha}}\left|\psi_{n_\alpha}^{(k)}\right\>\right)\otimes\left(\bigotimes_{\beta=M+1}^{M+N}\sqrt{{n_\beta+j-1\choose n_\beta}}\left|\psi_{n_\beta}^{(j)}\right\>\right)
\end{align*}
where $\{n_i\}:=\{n_1,n_2,\dots,n_{N+M}\}$ is a possible partition of $n$ into $N+M$ non-negative integers and the summation is over the set $\grp P_{n,N+M}$ consisting of all such partitions. We further note the orthonormality of $\left\{\left|\Psi_{M,N,n}^{(k,j)}\right\>\right\}_{n}$ so that Eq.~\eqref{Ncopy} is a proper expansion of $|\xi, k\>^{\otimes M}|\xi ,  j \>^{\otimes N}$. Notice that
\begin{align*}
\left\<\Psi_{M,N,n}^{(k,j)}\right|\left.\Psi_{M,N,m}^{(k,j)}\right\>&=\sqrt{{kM+jN+n-1\choose n}^{-1}{kM+jN+m-1\choose m}^{-1}}\times\\&\sum_{\{n_i\}\in\grp P_{n,(M+N)}}\sum_{\{m_i\}\in\grp P_{m,(M+N)}}\left[\prod_{\alpha=1}^{M}\sqrt{{n_\alpha+k-1\choose n_\alpha}{m_\alpha+k-1\choose m_\alpha}}\delta_{n_\alpha,m_\alpha}\right]\left[\prod_{\beta=M+1}^{M+N}\sqrt{{n_\beta+j-1\choose n_\beta}{m_\beta+j-1\choose m_\beta}}\delta_{n_\beta,m_\beta}\right]\\
&=\delta_{m,n}{kM+jN+n-1\choose n}^{-1}\sum_{\{n_i\}\in\grp P_{n,(M+N)}}\left[\prod_{\alpha=1}^{M}{n_\alpha+k-1\choose n_\alpha}\right]\left[\prod_{\beta=M+1}^{M+N}{n_\beta+j-1\choose n_\beta}\right]\\
&=\delta_{m,n}{kM+jN+n-1\choose n}^{-1}\sum_{\{n_i\}\in\grp P_{n,(M+N)}}\left[\prod_{\alpha=1}^{M}(-1)^{n_\alpha}{-k\choose n_\alpha}\right]\left[\prod_{\beta=M+1}^{M+N}(-1)^{n_\beta}{-j\choose n_\beta}\right]\\
&=\delta_{m,n}(-1)^{n}{kM+jN+n-1\choose n}^{-1}\sum_{\{n_i\}\in\grp P_{n,(M+N)}}\left[\prod_{\alpha=1}^{M}{-k\choose n_\alpha}\right]\left[\prod_{\beta=M+1}^{M+N}{-j\choose n_\beta}\right]\\
&=\delta_{m,n}(-1)^{n}{-kM-jN\choose n}{kM+jN+n-1\choose n}^{-1}=\delta_{m,n},
\end{align*}
having used the Chu-Vandermonde identity (see e.g. Chp. 7 of \cite{chu}).

With the proper expansion Eq.~\eqref{Ncopy} and the prior Eq.~\eqref{priorperelomov}, $\rho_\beta$ and $\Omega_\beta$ can be calculated. Notice that $|\xi , j\>^{\otimes N}$ can be expanded as
\begin{align*}
\rho_\beta&=\int p_\beta(\xi)~\mu(\d^2 \xi)~\left(|\xi, j\>\<\xi,  j|\right)^{\otimes N}\\
&=\sum_{m,n=0}^{\infty}\int_{0}^{2\pi}\frac{\d\theta}{2\pi}~e^{i(n-m)\theta}\int_{0}^{\infty}\frac{\d s~\beta(\sinh s)^{n+m+1}}{(\cosh s)^{2jN+\beta+n+m+1}}\sqrt{{jN+n-1\choose n}{jN+m-1\choose m}}\left|\Psi_{0,N,n}^{(k,j)}\right\>\left\<\Psi_{0,N,m}^{(k,j)}\right|\\
&=\sum_{n=0}^{\infty}\frac{\beta~{jN+n-1\choose n}}{(2jN+\beta){jN+\beta/2+n \choose n}}\left|\Psi_{0,N,n}^{(k,j)}\right\>\left\<\Psi_{0,N,n}^{(k,j)}\right|
\end{align*}
and we know that it is a diagonal form of $\rho_\beta$ from the orthogonality of $\left\{\left|\Psi_{0,N,n}^{(k,j)}\right\>\right\}_n$.
Similarly we have $\Omega_\beta$ diagonalized as
\begin{align*}
\Omega_\beta&=\int p_\beta(\xi)~\mu(\d^2 \xi)~\left(|\xi,  k\>\<\xi , k|\right)^{\otimes M} \otimes \left(|\xi , j\>\<\xi , j |\right)^{\otimes N} \\
&=\sum_{m=0}^{\infty}\frac{\beta~{kM+jN+m-1\choose m}}{(2kM+2jN+\beta){kM+jN+m+\beta/2 \choose m}}~\left|\Psi_{M,N,m}^{(k,j)}\right\>\left\<\Psi_{M,N,m}^{(k,j)}\right|.
\end{align*}

In the same spirit as Appendix~\ref{sec:proofqudit}, we now diagonalize $A_\beta:=\left(I\otimes\rho_\beta^{-\frac{1}{2}}\right)\Omega_\beta\left(I\otimes\rho_\beta^{-\frac{1}{2}}\right)$ by first defining
\begin{align*}
\left|\Phi_{M,N,m}^{(k,j)}\right\>:=\left(I\otimes\rho_\beta^{-\frac{1}{2}}\right)\left|\Psi_{M,N,m}^{(k,j)}\right\>
\end{align*}
so that
\begin{align}\label{diagAperelomov}
A_\beta=\sum_{m=0}^{\infty}\frac{\beta}{2kM+2jN+\beta}~{kM+jN+m-1\choose m}{kM+jN+m+\beta/2 \choose m}^{-1}~\left|\Phi_{M,N,m}^{(k,j)}\right\>\left\<\Phi_{M,N,m}^{(k,j)}\right|.
\end{align}
The vectors $\left|\Phi_{M,N,m}^{(k,j)}\right\>$ are mutually orthogonal. The proof can be completed by using the property
\begin{align*}
\left|\Psi_{M,N,m}^{(k,j)}\right\>=\sum_{n=0}^{m}\sqrt{{kM+m-n-1\choose m-n}{jN+n-1\choose n}{kM+jN+m-1\choose m}^{-1}}\left|\Psi_{M,0,m-n}^{(k,j)}\right\>\left|\Psi_{0,N,n}^{(k,j)}\right\>
\end{align*}
to calculate the inner product of any two vectors, obtaining
\begin{align*}
\left\<\Phi_{M,N,m}^{(k,j)}\right|\left.\Phi_{M,N,m'}^{(k,j)}\right\>&=\left\<\Psi_{M,N,m}^{(k,j)}\right|\left(I\otimes\rho_\beta^{-1}\right)\left|\Psi_{M,N,m'}^{(k,j)}\right\>\\
&=\sum_{n=0}^{m}\sum_{n'=0}^{m'}\sqrt{\frac{{kM+m-n-1\choose m-n}{jN+n-1\choose n}}{{kM+jN+m-1\choose m}}}\sqrt{\frac{{kM+m'-n'-1\choose m'-n'}{jN+n'-1\choose n'}}{{kM+jN+m'-1\choose m'}}}\left\<\Psi_{M,0,m-n}^{(k,j)}\right|\left.\Psi_{M,0,m'-n'}^{(k,j)}\right\>\left\<\Psi_{0,N,n}^{(k,j)}\right|~\rho_\beta^{-1}\left|\Psi_{0,N,n'}^{(k,j)}\right\>\\
&=\sum_{n=0}^{m}\sum_{n'=0}^{m'}\sqrt{\frac{{kM+m-n-1\choose m-n}{jN+n-1\choose n}}{{kM+jN+m-1\choose m}}}\sqrt{\frac{{kM+m'-n'-1\choose m'-n'}{jN+n'-1\choose n'}}{{kM+jN+m'-1\choose m'}}}\frac{(2jN+\beta){jN+\beta/2+n \choose n}}{\beta~{jN+n-1\choose n}}\delta_{m-n,m'-n'}\delta_{n,n'}\\
&=\delta_{m,m'}\sum_{n=0}^{m}\frac{(2jN+\beta){kM+m-n-1\choose m-n}{jN+\beta/2+n \choose n}}{\beta~{kM+jN+m-1\choose m}}\\
&=\delta_{m,m'}\frac{(2jN+\beta){kM+jN+\beta/2+m\choose m}}{\beta~{kM+jN+m-1\choose m}}.
\end{align*}
The last step comes from the properties of binomials and the Chu-Vandermonde identity \cite{chu}.

Finally, defining the normalized vectors $\left|\bar{\Phi}_{M,N,m}^{(k,j)}\right\>:=\frac{\left|\Phi_{M,N,m}^{(k,j)}\right\>}{\sqrt{\left\<\Phi_{M,N,m}^{(k,j)}\right|\left.\Phi_{M,N,m}^{(k,j)}\right\>}}$ [c.f. Appendix~\ref{sec:proofqudit}] and using Eq.~\eqref{diagAperelomov} we get the diagonalization
\begin{align*}
A_\beta=\frac{2jN+\beta}{2kM+2jN+\beta}\sum_{m=0}^{\infty}\left|\bar{\Phi}_{M,N,m}^{(k,j)}\right\>\left\<\bar{\Phi}_{M,N,m}^{(k,j)}\right|
\end{align*}
which is proportional to a projector. Consequently, the operator norm equals any one of the eigenvalues of $A_\beta$:
\begin{align*}
\|A_\beta\|_{\infty}=\frac{2jN+\beta}{2kM+2jN+\beta}.
\end{align*}
By Eq.~\eqref{aaa} we upper-bounded the MP fidelity $F^{(\rm P)}_{c}(\beta)\le(2jN+\beta)/(2kM+2jN+\beta)$. The upper bound is always achievable as shown in Eq. \eqref{probCFT}. Therefore we conclude that we have found the benchmark \eqref{eq:benchmarkperelomov}.

\subsection{Optimality of the square-root measurement in the case of uniform prior}\label{app:optsquare}

Consider the deterministic protocol that consists in measuring the square-root measurement   $\{  P_{\hat g}\}$ and re-preparing the generalized coherent state $|\psi_{\hat g}\>$.
   Its fidelity is given by
\begin{align*}
F_{\rm sqrt}   &    =
  \int  \d g  \,\int \d \hat g  \,     \<  \phi_g  |   P_{\hat g}  |\phi_{ g}\>   \,     |   \<  \psi_g  |  \psi_{\hat g}  \>|^2   \\
 &    =
  d_\phi  \int  \d g  \,\int \d \hat g  \,   |  \<  \phi_g  |\phi_{\hat g}\>|^2   \,     |   \<  \psi_g  |  \psi_{\hat g}  \>|^2   \\
&  =d_\phi \int  \d g  \,    | \<  \phi |\phi_g\> |^2  \,     |   \<  \psi |  \psi_{g}  \>|^2   \\
&  \equiv  F_{\rm  c}  \, .
\end{align*}
Here we used the definition of $P_{g}$ in Eq. (\ref{sqrtm}) in the first equality, the invariance of the Haar measure in the second equality, and  Eq.  (\ref{bench}) in the third equality.
In conclusion, the fidelity of the simple protocol based on the square-root measurement coincides with the ultimate probabilistic CFT, implying that, for uniform prior, there is no advantage from using a probabilistic strategy.

\subsection{Teleportation fidelity achieved by  the optimal square-root measurement}\label{sec:proofsquareroot}
Here we investigate the performance of the deterministic MP protocol that  consists in testing the input with the square-root measurement $\{P_\eta(g)\}_{g\in\grp{SU(2)}}$ [c.f. Eq.~\eqref{sqrtPOVM}] and preparing the output state that corresponds to the outcome.

As shown in Eq.~\eqref{sqrtfidelity}, the teleportation fidelity can be expressed as
\begin{align}\label{sqrtfidelity2}
F(\beta,\eta)=\Tr\left[\Omega_\beta(I\otimes\rho_\eta^{-\frac{1}{2}})\Omega_\eta(I\otimes\rho_\eta^{-\frac{1}{2}})\right] \, ,
\end{align}
where
\begin{align*}
\Omega_\eta&:=\int\d g~p_\eta(g)~ |\psi_g\>\!\<\psi_g|^{\otimes 2}\\
&=\frac{\eta+1}{\eta+3}\left\<\frac{\eta}{2}, \frac{\eta}{2}\right|P_{\eta/2+1}\left|\frac{\eta}{2}, \frac{\eta}{2}\right\>
\end{align*}
and
\begin{align*}
\rho_\eta&:=\int\d g~p_\eta(g)~ |\psi_g\>\!\<\psi_g|\\
&=\frac{\eta+1}{\eta+2}\left\<\frac{\eta}{2}, \frac{\eta}{2}\right|P_{\eta/2+1/2}\left|\frac{\eta}{2}, \frac{\eta}{2}\right\>.
\end{align*}
Here $|J, m\>$ denotes as usual the eigenstate corresponding to the eigenvalue $m$  of the $z$-component of the total angular momentum for a system of total spin $J$ and $P_{J}  :  = \sum_{m=-J/2}^{J/2}|J, m\>\<J, m|$.
Using the definition and the Clebsch-Gordan coefficients for the coupling of angular momenta, it is easy to obtain the relations
\begin{align*}
\rho_\eta=\frac{\eta+1}{\eta+2}\left|\frac{1}{2}, \frac{1}{2}\right\>\left\<\frac{1}{2}, \frac{1}{2}\right|+\frac{1}{\eta+2}\left|\frac{1}{2}, -\frac{1}{2}\right\>\left\<\frac{1}{2}, -\frac{1}{2}\right|
\end{align*}
and
\begin{align*}
\Omega_\eta=\frac{\eta+1}{\beta+3}|1, 1\>\<1, 1|+\frac{2(\eta+1)}{(\eta+3)(\beta+2)}|1, 0\>\<1, 0|+\frac{2}{(\eta+3)(\eta+2)}|1, -1\>\<1, -1|.
\end{align*}
Inserting these relations into Eq.~\eqref{sqrtfidelity2} we then obtain the fidelity
\begin{align*}
F(\beta,\eta)=\frac{\eta+2}{\eta+3}\cdot\frac{\beta+1}{\beta+3}+\frac{(1+\sqrt{\eta+1})^2}{\eta+3}\cdot\frac{\beta+1}{(\beta+2)(\beta+3)}+\frac{4}{(\eta+3)(\beta+3)(\beta+2)} \, ,
\end{align*}
which, optimized over $\eta$, yields  the value
\begin{align}
\label{sqrtopt}
\max_{\eta}F(\beta,\eta)     =   F_{\rm c}^{(2){\rm tele}}(\beta)  -  \frac{4\beta}{(\beta+2)(\beta+3)\left[(\beta+1)(\beta+3)+\sqrt{\beta^4+8\beta^3+22\beta^2+8\beta^2+9}\right]} \, ,
\end{align}
where $F_{\rm c}^{(2){\rm tele}}(\beta)=\frac{\beta+2}{\beta+3}$ is the teleportation benchmark of Eq.~\eqref{qubittele}.

Interestingly, the optimal value of $\eta$, given by
\begin{align*}
\eta_{\rm opt}=\frac{\beta^4+8\beta^3+16\beta^2-4\beta+3+(\beta^2+4\beta-1)\sqrt{\beta^4+8\beta^3+22\beta^2+8\beta+9}}{2(\beta+2)^2},
\end{align*}
is not equal to $\beta$.
In summary, \emph{i)} our parametric family of square-root measurements does not achieve the probabilistic benchmark for any finite value of $\beta$ and \emph{ii)} the square-root measurement associated to the input ensemble is \emph{not optimal} among the measurements in this parametric family.

\subsection{Proof of the optimality of MP protocols for the complex conjugation of generalized coherent states}\label{sec:cc}
Here we prove that there is no quantum advantage in the  complex conjugation of GPCS, i.e.~for the task of transforming $N$ copies of the input GPCS  $|\phi_g\>$, given  with  prior probability  $p_\gamma(g)$ as in Eq.~(\ref{prior}), into $M$ copies of the output  GPCS $|\psi^*_g\>$.

For the complex conjugation of GPCS, the optimal quantum fidelity can be obtained from Eq.~(\ref{qfid}), which now reads
\begin{align*}
F_{\rm q}(\gamma)   =  \left\| \left(  I_{\rm out}  \otimes   \rho_\gamma^{-\frac 12}\right)     \Omega_\gamma   \left(  I_{\rm out}  \otimes   \rho_\gamma^{-\frac 12}\right) \right\|_\infty   \equiv \lambda_{\min} \, ,
\end{align*}
where the operators $\rho_\gamma$ and $\Omega_\gamma$ are given by
\begin{align}
\rho_\gamma  &:=  \int  \d g \,    p_\gamma (g) \,  |\phi_g\>\< \phi_g|  \\
\Omega_\gamma  &   :=    \int  \d g \,    p_\gamma (g) \,  |\psi_g\>\<  \psi_g|  \otimes |\phi_g\>\< \phi_g|  \, .
\end{align}
The two operators $\rho_\gamma$ and $\Omega_\gamma$ defined here coincide with the operators   $\rho_\gamma$ and $\Omega_\gamma$ in the previous section of the Appendix, where we have shown an upper bound on  the operator norm.  Replacing $\Omega_\gamma$ by $\Omega^{T_{\rm out}}_\gamma$, the  upper bound of Eq.~(\ref{uppe}) now reads
\begin{align}
\left\| \left(  I_{\rm out}  \otimes   \rho_\gamma^{-\frac 12}\right)     \Omega_\gamma   \left(  I_{\rm out}  \otimes   \rho_\gamma^{-\frac 12}\right) \right\|_\infty    \le  \frac{\int \d g  \,       |\< \psi|\psi_g\> |^{2 M}        |\< \phi|\phi_g\> |^{2 N}   |\<  \phi_\gamma|\phi_{\gamma,g}\>|^2}{\int \d g  \,           |\< \phi|\phi_g\> |^{2 N}   |\<  \phi_\gamma|\phi_{\gamma,g}\>|^2} \equiv  F_{\rm c}  (\gamma)  \, .
\end{align}
Hence, the ultimate quantum fidelity is upper bounded by the CFT.  Since by definition
 $F_{\rm q}  (\gamma) \ge F_{\rm c}  (\gamma)$, this means that one has the equality $F_{\rm q}  (\gamma)  = F_{\rm c}  (\gamma)$.    In other words, MP protocols   and general quantum strategies fare equally well.



\end{widetext}

\end{document}